\documentclass[aps,prx,fixfloat,twocolumn]{revtex4-2}
\usepackage{graphicx,amsmath,amsfonts,bm, color}
\usepackage{wasysym}
\usepackage{xcolor,hyperref}
\hypersetup{
   colorlinks,
   linkcolor={blue!50!black},%{red!80!black},
   citecolor={blue!50!black},
   urlcolor={blue!80!black}
}
\usepackage{enumitem}

\renewcommand{\=}{\!=\!}
\newcommand{\1}{^{\mbox{\tiny (1)}}}

\DeclareMathAlphabet{\mathitbf}{OML}{cmm}{b}{it}

\begin{document}

\title{Facet formation in slow three-dimensional fracture}
\author{Yuri Lubomirsky}
\author{Eran Bouchbinder}
%\email{eran.bouchbinder@weizmann.ac.il}
\affiliation{Chemical and Biological Physics Department, Weizmann Institute of Science, Rehovot 7610001, Israel}

\begin{abstract}
Cracks develop various surface patterns as they propagate in three-dimensional (3D) materials. Facet formation in nominally tensile (mode-I) fracture emerge in the slow, non-inertial regime and oftentimes takes the form of surface steps. We show that the same phase-field framework that recently shed basic light on dynamic (inertial) tensile fracture in 3D, also gives rise to crack surface steps. Step formation is shown to be an intrinsically nonlinear phenomenon that involves two essential physical ingredients: finite-strength quenched disorder and a small, mesoscopic anti-plane shear (mode-III) loading component (on top of the dominant tensile, mode-I loading component). We quantify the interplay between disorder (both its strength and spatial correlation length) and mesoscopic mode I+III mixity in controlling step formation. Finally, we show that surface steps grow out of the small-scale, background surface roughness and are composed of two overlapping crack segments connected by a bridging crack, in agreement with experiments.
\end{abstract}

\maketitle

{\em Introduction}.---Crack propagation is the main process that mediates material failure. Yet, despite its prime scientific and technological importance, our understanding of the spatiotemporal dynamics of cracks is still incomplete, especially in 3D. In 3D fracture, under nominally tensile (mode-I) conditions, cracks are known to form various surface structures out of the tensile symmetry plane~\cite{gent1984micromechanics,tanaka1998discontinuous,tanaka2000fracture,baumberger2008magic,ronsin2014crack,steps2017,steinhardt2022material,wang2022hidden,steinhardt2023geometric,wang2023dynamics,pham2014further,ravi1984experimental_II,ravi1984experimental_III,scheibert2010brittle,guerra2012understanding,sharon1996microbranching,sharon1998universal,sharon1999dynamics,livne2005universality,lawn,johnson1968microstructure,rabinovitch2000origin,jiao2015macroscopic,lubomirsky2023quenched}. In the slow fracture regime, i.e., when the crack propagation velocity is much smaller than elastic wave-speeds, faceted fracture surfaces emerge, oftentimes featuring symmetry-breaking topological defects in the form of surface steps~\cite{gent1984micromechanics,tanaka1998discontinuous,tanaka2000fracture,baumberger2008magic,ronsin2014crack,steps2017,steinhardt2022material,wang2022hidden,steinhardt2023geometric,wang2023dynamics,pham2014further}.

Step formation is not yet fully understood. In particular, linear stability analyses~\cite{ball1995three,movchan1998perturbations,leblond2011theoretical,leblond2016out} and recent computer simulations~\cite{lebihain2020effective,lubomirsky2023quenched} show that slow cracks in 3D are linearly stable against out-of-plane perturbations under pure tensile loading conditions. These results suggest that basic physical ingredients might be missing in our current understanding of step formation. One possible ingredient might be related to nonlinear effects~\cite{ronsin2014crack,leblond2016out,chen2015}, in particular finite quenched disorder (characterized by finite strength and correlation length), which has been very recently shown to play decisive roles in dynamic (inertial) tensile fracture in 3D~\cite{lubomirsky2023quenched}.

\begin{figure}[ht!]
\center
\includegraphics[width=0.38\textwidth]{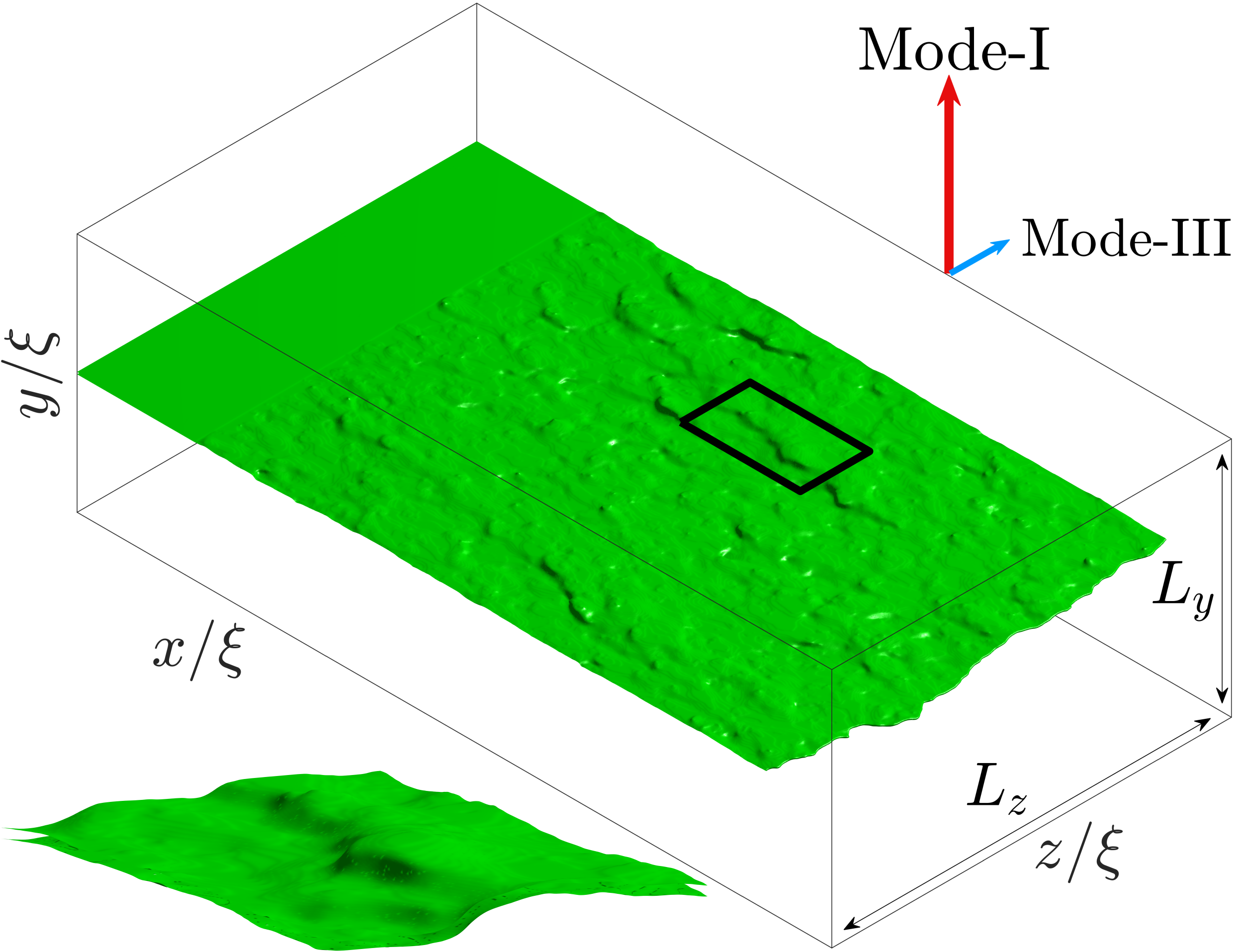}
\vspace{-0.25cm}
\caption{The fracture surface (green) generated by a crack in a long bar of height $L_y$ and thickness $L_z$ under predominantly tensile (mode-I, long red arrow) loading. A small anti-plane shear (mode-III, short blue arrow) loading component is superimposed. The initial flat notch is located in the middle plane and the crack subsequently propagated predominantly in the $x$ direction. The coordinate system $(x,y,z)$ is marked (in units of the dissipation length $\xi$), and the parameters used are $K_{_{\rm III}}/K_{_{\rm I}}\!=\!0.05$, $\sigma\!=\!0.5$ and $R/\xi\!=\!5$, see text for details. (inset) A zoom-in on a ridge line (see text) corresponding to the rectangle in the main panel.}
\vspace{-0.3cm}
\label{fig:fig1}
\end{figure}

Another possible missing ingredient has been suggested by recent experiments on brittle polymer gels~\cite{wang2022hidden}. It has been shown that experiments on nominally tensile (mode-I) fracture quite often include a small, uncontrolled anti-plane shear (mode-III) loading component. Importantly, it has been demonstrated that once the small mode-III loading component is carefully eliminated, surface facets disappear, i.e., these results indicate that step formation crucially depends on the existence of a small mode I+III mixity (see Fig.~\ref{fig:fig1}).
%%%%%%%%%%%%%%%%%%%%%%%%%%%%%%%%%%%%%%%
%%%%%%%%%%%%%%%%%%%%%%%%%%%%%%%%%%%%%%%
%%%%%%%%%%%%%%%%%%%%%%%%%%%%%%%%%%%%%%%
\begin{figure*}[ht!]
\center
\includegraphics[width=0.95\textwidth]{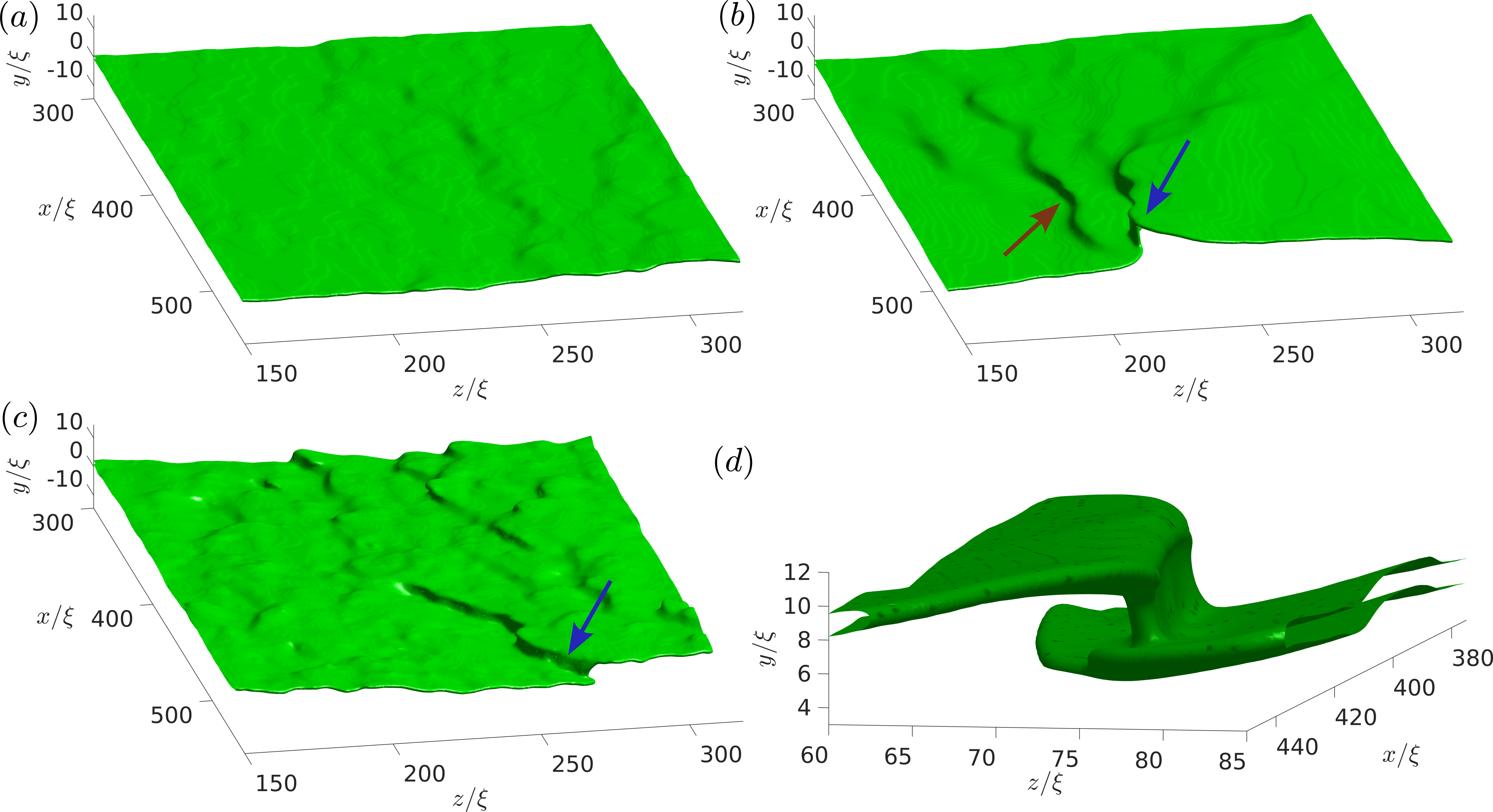}
\vspace{-0.25cm}
\caption{(a) A fracture surface generated by a 3D crack with $K_{_{\rm III}}/K_{_{\rm I}}\!=\!0.15$, $\sigma\!=\!0.25$ and $R/\xi\!=\!5$. (b) Same as panel (a), but with $R/\xi\!=\!10$. (c) Same as panel (a), but with $\sigma\!=\!0.5$. (d) A zoom-in on a step generated with $K_{_{\rm III}}/K_{_{\rm I}}\!=\!0.2$, $\sigma\!=\!0.5$ and $R/\xi\!=\!10$ (there is another, coexisting step, which is not shown). See text for discussion.}
\vspace{-0.3cm}
\label{fig:fig2}
\end{figure*}
%%%%%%%%%%%%%%%%%%%%%%%%%%%%%%%%%%%%%%%

These observations are in line with earlier experiments on soft hydrogels~\cite{ronsin2014crack}, where steps/facets observed in the presence of a controlled, global mode I+III mixity --- similarly to a large body of literature on mixed-mode I+III fracture~\cite{pollard1982formation,lazarus2008comparison,pons2010helical,leblond2011theoretical,goldstein2012fracture,leblond2015multiscale,chen2015,leblond2019configurational,vasudevan2020configurational,lin2010criterion,pham2016growth,pham2017formation,leblond2016out} --- were related to steps that emerged in nominally mode-I fracture~\cite{baumberger2008magic}. The latter steps were attributed to sufficiently large mesoscopic structural fluctuations accompanied by some mode I+III mixity, and as such precisely highlight the two physical ingredients discussed above. This picture is also consistent with experiments on glasses and brittle thermosetting polymers~\cite{sommer1969formation,pham2014further}, where steps emerged under very small levels of controlled, global mode I+III mixity. Finally, hydraulic fracture experiments on brittle hydrogels demonstrated the effect of material heterogeneity on step formation through the addition of microbeads of various sizes and number density~\cite{steinhardt2022material}.

Overall, our proposed physical picture in which steps emerge from a combination of a small, mesoscopic mode I+III mixity and finite-strength quenched disorder appears to be consistent with experiments on a broad range of amorphous (disordered) materials. Yet, it has never been directly and conclusively demonstrated. Moreover, basic questions regarding step nucleation and subsequent spatiotemporal dynamics remain open. Here, we study slow 3D fracture using a flexible computational framework in which material quenched disorder is incorporated. It is based on a phase-field fracture approach~\cite{lubomirsky2018,chen2017,vasudevan2021oscillatory,karma2001phase,Karma2004,Hakim.09,das2023dynamics,bleyer2017microbranching,henry2013fractographic,lubomirsky2023quenched}, which has been very recently shown to unprecedentedly predict both the oscillatory and branching instabilities in 2D dynamic fracture~\cite{lubomirsky2018,chen2017,vasudevan2021oscillatory}, and the inertial dynamics of tensile fracture in 3D~\cite{das2023dynamics,lubomirsky2023quenched}.

We show that step formation is an intrinsically nonlinear phenomenon that crucially involves {\em both} of the two aforementioned physical ingredients: quenched disorder (of finite strength and correlation length) and a small, mesoscopic anti-plane shear (mode-III) loading component. We numerically compute a comprehensive phase diagram for step formation in terms of quenched disorder and mode I+III mixity. We also show that steps grow out of background surface roughness ridges and are composed of two overlapping crack segments connected by a bridging crack, in line with experiments.

{\em The emergence of steps in slow 3D fracture}.---We employ the recently-developed, disordered 3D phase-field fracture framework of~\cite{lubomirsky2023quenched}, where displacement field $\bm{u}({\bm x},t)$ is coupled to an auxiliary field, the scalar phase-field $\phi({\bm x},t)$~\cite{lubomirsky2018,chen2017,vasudevan2021oscillatory,das2023dynamics,SI,lubomirsky2023quenched}. The latter satisfies its own dissipative field equation, featuring a characteristic dissipation length $\xi$ and a dissipation time~\cite{lubomirsky2018,chen2017,vasudevan2021oscillatory,das2023dynamics,SI,lubomirsky2023quenched}, which manifest themselves near a crack front (cf.~Fig.~\ref{fig:fig1}, also for the definition of the Cartesian coordinate system ${\bm x}\=(x,y,z)$, and note that $t$ is time).

Over spatial scales larger than $\xi$ away from the front, one has $\phi({\bm x},t)\=1$ and $\bm{u}({\bm x},t)$ corresponds to linear elasticity, featuring a nearly singular gradient (${\bm\nabla}\bm{u}\!\sim\!1/\sqrt{r}$, where $r$ is the distance from the front~\cite{Freund,lawn}). For $r\!\lesssim \!\xi$, $\phi({\bm x},t)$ drops towards zero, spontaneously generating the traction-free boundary conditions that define the crack and giving rise to a rate-dependent fracture energy $\Gamma(v)$ ($v$ is the crack propagation velocity). The crack trajectory is spontaneously selected without invoking any extraneous path-selection criteria, allowing to track the in silico real-time 3D spatiotemporal crack dynamics in a way that goes well beyond current experiments. The crack is defined as the $\phi({\bm x},t)\!=\!1/2$ iso-surface.

We employ the same disordered 3D phase-field fracture model of~\cite{lubomirsky2023quenched}, including the sample geometry, numerical implementation of the field equations and large-scale GPU-based computer simulations~\cite{lubomirsky2023quenched,SI}. Quenched disorder, which is a generic feature of amorphous materials, is incorporated into the static fracture energy $\Gamma_0\=\Gamma(v\=0)$ such that $\Gamma({\bm x};v\=0)/\Gamma_0$ is a Gaussian field characterized by unity mean, width $\sigma$ and spatial correlation length $R$~\cite{lubomirsky2023quenched,SI}. It has been recently shown to play crucial roles in predicting the 3D dynamics of tensile (mode-I) cracks in the inertial regime (i.e., where $v$ is comparable to elastic wave-speeds), in agreement with a broad range of experimental observations~\cite{lubomirsky2023quenched}

There are only two differences compared to~\cite{lubomirsky2023quenched}. First, as step formation is experimentally observed in slow 3D fracture, we focus on the quasi-static regime~\cite{SI}, where $v$ is much smaller than elastic wave-speeds. Second, we incorporate a small mode-III loading component in addition to the dominant mode-I one, quantified by the mode I+III mixity $K_{_{\rm III}}/K_{_{\rm I}}$~\cite{SI}. Here, $K_{_{\rm I}}$ and $K_{_{\rm III}}$ are the mode-I and mode-III stress intensity factors (SIFs), respectively (the SIF is the amplitude of the linear elastic $1/\sqrt{r}$ stress singularity for each symmetry mode~\cite{Freund,lawn}). We fix the dimensionless crack driving force $G/\Gamma_0$, where $G\=\left[(1-\nu)K_{_{\rm I}}^2+K_{_{\rm III}}^2\right]\!/2\mu$ is the energy release rate~\cite{Freund,lawn}. Finally, we employ periodic boundary conditions across the thickness direction (of size $L_z$, see~Fig.~\ref{fig:fig1}), and set $L_y\=192\xi$ and $L_z\=318\xi$. The latter value, which is much smaller compared to macroscopic $L_z$ values, implies that mode I+III mixity in our computations represents {\em mesoscopic} mode-mixity, which may arise from a variety of experimental and/or material sources.

We study the emergence of out-of-plane crack structures in slow 3D fracture as a function of $K_{_{\rm III}}/K_{_{\rm I}}\!\ll\!1$ and the quenched disorder parameters $\sigma$ and $R/\xi$. In Fig.~\ref{fig:fig2}a, we present a fracture surface generated by a crack with $K_{_{\rm III}}/K_{_{\rm I}}\=0.15$, $\sigma\=0.25$ and $R/\xi\=5$. Here, the surface features out-of-plane roughness, but a distinct step does not emerge above the small-scale roughness level. Out-of-plane roughness has been extensively discussed in the literature, mainly in the context of self-affine scaling properties (e.g.,~\cite{bonamy2011failure}), but is not commonly studied along with step formation. Notable exceptions are~\cite{tanaka2000fracture,ronsin2014crack}, where small-scale roughness has been demonstrated both in the absence of steps and in their presence. Specifically in~\cite{baumberger2008magic,ronsin2014crack}, roughness has been shown to reveal ridges in the crack propagation direction (see Fig.~2b in~\cite{baumberger2008magic}) and when steps emerged, they appeared to grow out of these roughness ridges~\cite{ronsin2014crack}.

Shallow roughness ridges are observed in Fig.~\ref{fig:fig2}a (e.g., the ridge line along the propagation direction $x$, in the vicinity of $z/\xi\!\simeq\!250$). Ridge lines are localized out-of-plane front distortions that persist mainly in the crack propagation direction. Their out-of-plane distortion angle is smaller than 90$^{\circ}$, leaving the front continuous (unlike steps, which involve a topological change). Steeper ridge lines are observed in Fig.~\ref{fig:fig1}, where the inset (bottom-left corner) shows a zoom-in on part of the ridge line marked by the black rectangle in the main panel. Ridges, and the distinction between them and steps, are also discussed in relation to Fig.~\ref{fig:fig5}.

In Fig.~\ref{fig:fig2}b, we present part of a fracture surface emerging in a simulation as in Fig.~\ref{fig:fig2}a, with everything being the same except that we set $R/\xi\=10$. A distinct step (marked by the blue arrow) is observed, involving a 90$^{\circ}$ facet that is followed by a topological change, resulting in two crack segments (as will be further discussed in relation to Fig.~\ref{fig:fig2}d and Fig.~\ref{fig:fig5}). The step in Fig.~\ref{fig:fig2}b appears to grow out of a roughness ridge. A coexisting ridge, which eventually did not transform into a step, is marked by the red arrow.

In Fig.~\ref{fig:fig2}c, we present part of a fracture surface emerging in a simulation as in Fig.~\ref{fig:fig2}a, with everything being the same expect that we set $\sigma\=0.5$. The larger amplitude disorder results in steeper multiple ridges, one of which transforms into a step (marked by the blue arrow) that drifts at an angle relative to the crack propagation direction. These highly localized steps are consistent with corresponding observations in a broad range of experiments~\cite{gent1984micromechanics,tanaka1998discontinuous,tanaka2000fracture,baumberger2008magic,ronsin2014crack,steps2017,steinhardt2022material,wang2022hidden,steinhardt2023geometric,wang2023dynamics,pham2014further}. Figure~\ref{fig:fig2} constitutes a major finding that steps formation in slow 3D fracture involves a small mesoscopic mode I+III mixity and finite-strength quenched disorder, of either sufficiently large correlation length $R$ or sufficiently large amplitude $\sigma$. Our next goals are to quantify the onset conditions for steps, and to gain deeper insight into their dynamics and topology/geometry.

\begin{figure}[ht!]
\center
\includegraphics[width=0.485\textwidth]{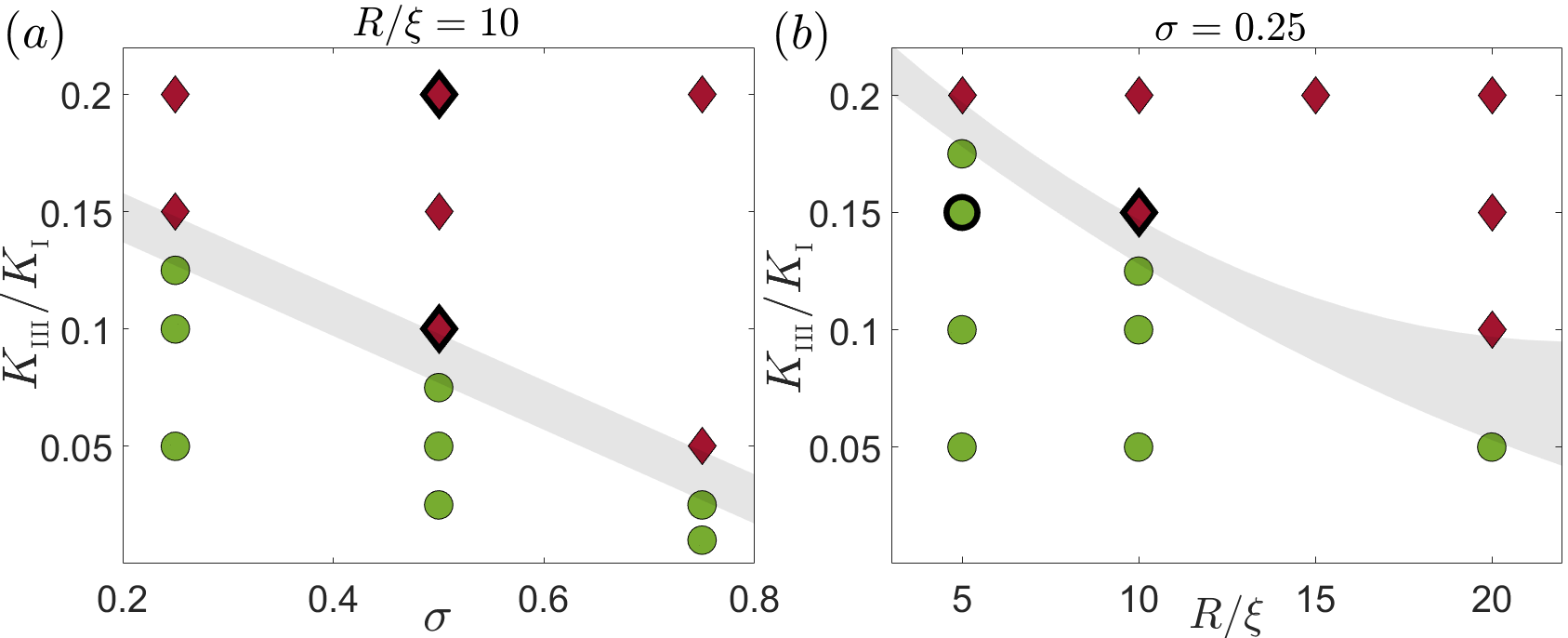}
\vspace{-0.25cm}
\caption{(a) Step formation phase diagram in the $K_{_{\rm III}}/K_{_{\rm I}}\!-\!\sigma$ plane for fixed $R/\xi\!=\!10$. Diamonds correspond to steps and circles for their absence, and the shaded area marks the transition region (guide to the eye). (b) Same as panel (a), but in the $K_{_{\rm III}}/K_{_{\rm I}}\!-\!R/\xi$ plane for fixed $\sigma\!=\!0.25$. Spatial patterns that correspond to the thick-boundary symbols appear in either Fig.~\ref{fig:fig2} or Fig.~\ref{fig:fig5}.}
\vspace{-0.3cm}
\label{fig:fig3}
\end{figure}

{\em Step formation phase diagram}.---The above results indicate that step formation depends on $K_{_{\rm III}}/K_{_{\rm I}}$, $\sigma$ and $R/\xi$, and that the threshold in $K_{_{\rm III}}/K_{_{\rm I}}$ is a decreasing function of both $\sigma$ and $R/\xi$. To quantify the onset conditions for steps, we present in Fig.~\ref{fig:fig3} two planar cuts of the step formation phase diagram in the $K_{_{\rm III}}/K_{_{\rm I}}\!-\!\sigma\!-\!R/\xi$ space.
%%%%%%%%%%%%%%%%%%%%%%%%%%%%%%%%%%%%%%%
%%%%%%%%%%%%%%%%%%%%%%%%%%%%%%%%%%%%%%%

We observe that indeed the mode I+III mixity threshold for step formation is a decreasing function of $\sigma$ (Fig.~\ref{fig:fig3}a) and $R/\xi$ (Fig.~\ref{fig:fig3}b). Moreover, the threshold becomes small for large $\sigma$ and/or $R/\xi$, which is consistent with experimental observations regarding a very small mode I+III mixity threshold for facet formation and hence indirectly supports the importance of finite-strength disorder in the amorphous materials employed. The finite values of $\sigma$ and $R/\xi$, and the transition triggered by increasing them, also highlight the intrinsically nonlinear, disorder-induced nature of step formation.

We note that while the phase diagram in Fig.~\ref{fig:fig3} appears to be binary (i.e., either steps emerge or not), the disorder-induced nature of step formation indicates that it is in fact a probabilistic process. Consequently, the actual transition may be smoother and likely to be better quantified in terms of probabilities. Indeed, in hydraulic fracture experiments on brittle hydrogels in which discrete material heterogeneity was externally controlled through the addition of microbeads~\cite{steinhardt2022material}, it has been demonstrated that the step formation probability increases with the microbeads size --- that is directly analogous to the correlation length $R$ ---, as well as with their number density.

Overall, the crucial role of material quenched disorder in step formation in slow 3D fracture seems to bear close analogy to its roles in relation to the localized branching typically observed at higher crack propagation velocities, including in the inertial regime, as discussed extensively in~\cite{lubomirsky2023quenched}. As such, quenched disorder --- and its associated lengthscale $R$ --- appears to be an essential physical ingredient in 3D fracture of amorphous materials.

\begin{figure}[ht!]
\center
\includegraphics[width=0.45\textwidth]{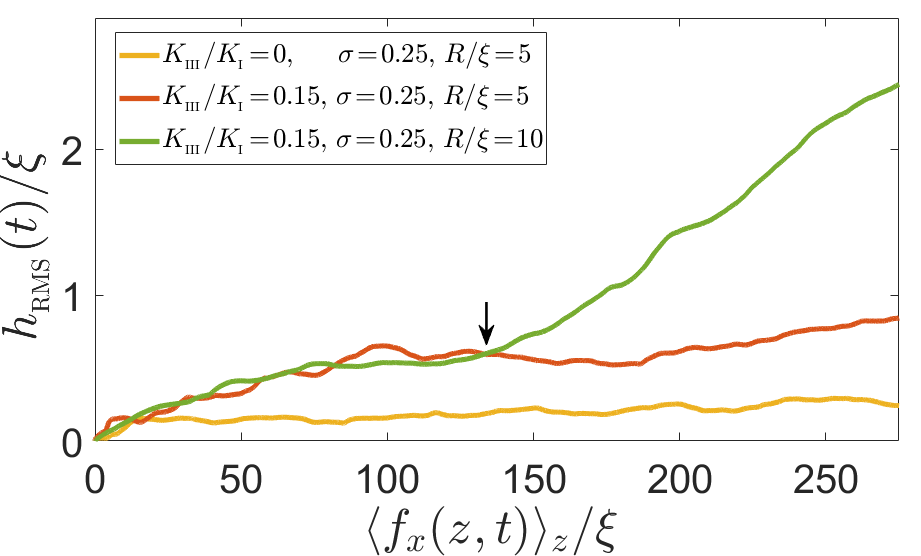}
\vspace{-0.25cm}
\caption{The RMS roughness $h_{\mbox{\tiny RMS}}(t)/\xi$ vs.~$\langle f_x(z,t)\rangle_z$ for the parameters indicated in the legend. See text for discussion.}
\vspace{-0.3cm}
\label{fig:fig4}
\end{figure}

\begin{figure}[ht!]
\center
\includegraphics[width=0.4\textwidth]{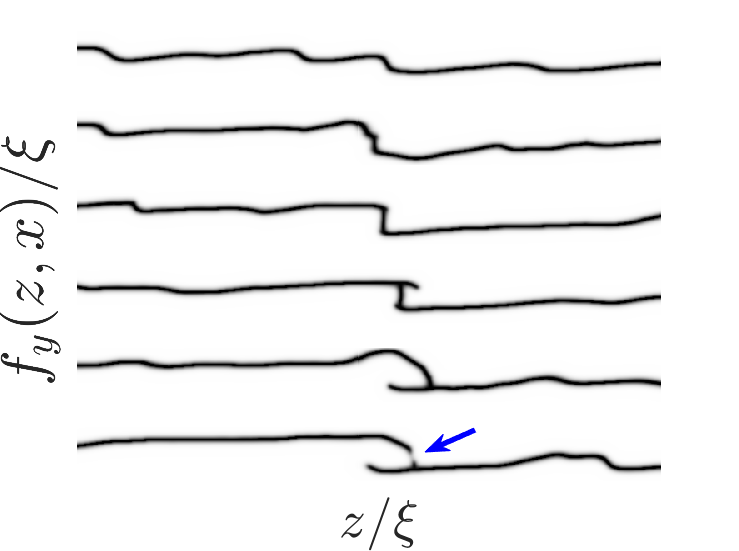}
\vspace{-0.25cm}
\caption{A sequence of $f_y(z,x)$ profiles corresponding to $K_{_{\rm III}}/K_{_{\rm I}}\!=\!0.1$, $\sigma\!=\!0.5$ and $R/\xi\!=\!10$. The profiles (from top to bottom) are shifted vertically for visual clarity and the line thickness corresponds to the variation of $\phi$ (from 1 to 0).}
\vspace{-0.3cm}
\label{fig:fig5}
\end{figure}

{\em Small-scale roughness, spatiotemporal facet dynamics and step topology/geometry}.---To quantify the spatiotemporal dynamics of fracture facets, we start by plotting in Fig.~\ref{fig:fig4} the out-of-plane root-mean-square (RMS) roughness $h_{\mbox{\tiny RMS}}(t)\=\sqrt{L_z^{-1}\!\!\int_{0}^{L_z}\![f_y(z,t)-\langle f_y(z,t)\rangle_z]^2\,dz}$ vs.~$\langle f_x(z,t)\rangle_z$ of several crack surfaces. Here, ${\bm f}(z,t)\=(f_x(z,t),f_y(z,t))$ is the front position and $\langle\,\cdot\,\rangle_z$ is an average over the thickness dimension $z$. The lowest curve corresponds to a pure mode-I crack ($K_{_{\rm III}}/K_{_{\rm I}}\!=\!0$), with $\sigma\!=\!0.25$ and $R/\xi\!=\!5$. The roughness level, $h_{\mbox{\tiny RMS}}\!\simeq\!0.2\xi$, is much smaller than $R$, demonstrating the large resistance of the crack front to out-of-plane distortion. The spatial distribution of roughness (not shown) is anisotropic, revealing persistent ridges in the crack propagation direction, as shown in Fig.~\ref{fig:fig2} and as observed experimentally~\cite{ronsin2014crack}.
%%%%%%%%%%%%%%%%%%%%%%%%%%%%%%%%%%%%%%%
%%%%%%%%%%%%%%%%%%%%%%%%%%%%%%%%%%%%%%%

One expects that the introduction of a small mode-III loading component would give rise to larger ridges and overall roughness. Indeed, setting $K_{_{\rm III}}/K_{_{\rm I}}\!=\!0.15$ --- corresponding to the fracture surface shown in Fig.~\ref{fig:fig2}a ---, larger ridges are observed and the roughness level increases, see orange curve in Fig.~\ref{fig:fig4}, yet no step emerges. Increasing the correlation length $R$ at fixed $K_{_{\rm III}}/K_{_{\rm I}}$ and $\sigma$ is expected to leave the roughness level roughly unchanged, but can lead to step formation. Consequently, we expect that setting $R/\xi\!=\!10$, such that we have $K_{_{\rm III}}/K_{_{\rm I}}\!=\!0.15$, $\sigma\!=\!0.25$ and $R/\xi\!=\!10$ as in Fig.~\ref{fig:fig2}b, $h_{\mbox{\tiny RMS}}(t)$ would overlap the roughness of the $R/\xi\!=\!5$ case (orange curve in Fig.~\ref{fig:fig4}) for some time, but then significantly depart from it once a step forms. This is indeed observed, see green curve in Fig.~\ref{fig:fig4} and the arrow that marks step formation.

Our main finding that steps can form upon increasing either $\sigma$ or $R$ at fixed small, mesoscopic $K_{_{\rm III}}/K_{_{\rm I}}$, together with the indications that roughness ridges may serve as nucleation sites for steps, indicate that out-of-plane distortions of the continuous crack front should be strong enough, either in magnitude and/or in angle, for steps to emerge. Moreover, the subsequent growth and eventually stabilization of steps imply that nontrivial crack structures should emerge, as described in~\cite{tanaka1998discontinuous,steps2017,wang2022hidden}, preventing the step from decaying back to the roughness level. To address this important issue, we present in Fig.~\ref{fig:fig2}d a zoom-in on one of the steps emerging in our calculations (see figure caption).

It is observed that the upper and lower parts of a distorted front form two crack segments (i.e., undergo a topological change~\cite{tanaka1998discontinuous,steps2017,wang2022hidden}) that reside at different overlapping $y$ planes, which eventually connect through a predominantly vertical bridging crack~\cite{pham2014further}. This topological/geometric structure, composed of several interacting cracks, appears to produce an effective repulsion between the overlapping segments, as envisioned and stressed in~\cite{steps2017}. Figure~\ref{fig:fig2}d further highlights the 3D multi-crack nature of surface steps~\cite{tanaka1998discontinuous,steps2017,wang2022hidden}.
%%%%%%%%%%%%%%%%%%%%%%%%%%%%%%%%%%%%%%%
%%%%%%%%%%%%%%%%%%%%%%%%%%%%%%%%%%%%%%%

Finally, to further elucidate the spatiotemporal dynamics of step nucleation and subsequent evolution --- complementing Fig.~\ref{fig:fig2}d ---, we present in Fig.~\ref{fig:fig5} a sequence of $f_y(z,x)$ profiles (the out-of-plane front at fixed $x$ values) before, during and after step nucleation. The first (uppermost) snapshot shows the background roughness, featuring a few ridges. One of the ridges, which we choose to locate in the middle, strongly rotates out of the plane and grows in amplitude (two subsequent snapshots). The forming step then transforms into two overlapping, nearly parallel crack segments, connected by a predominantly vertical bridging crack~\cite{pham2014further}. Subsequently, one crack segment curves down and intersects with the other segment, leaving another ``hidden'' segment beneath it and forming a ``$\tau$-structure''~\cite{tanaka1998discontinuous,steps2017,wang2022hidden}.

In the last snapshot (lowermost), an unbroken ligament starts to emerge (marked by the blue arrow) between two ``hand-shaking'' segments~\cite{pham2014further}, which will eventually fail under the intense stress fields that develop. The results in Fig.~\ref{fig:fig5} demonstrate how steps form out of surface roughness ridges and the subsequent evolution of their topology/geometry. They agree with a broad range of direct and indirect experimental observations~\cite{gent1984micromechanics,tanaka1998discontinuous,tanaka2000fracture,baumberger2008magic,ronsin2014crack,steps2017,steinhardt2022material,wang2022hidden,steinhardt2023geometric,wang2023dynamics,pham2014further}, but also go beyond experiments in revealing all stages of step formation in slow 3D fracture.

{\em Summary and outlook}.---Our results show that step formation in slow 3D fracture is a nonlinear, highly localized process that depends on both a small, mesoscopic mode I+III mixity and quenched disorder, of finite amplitude and correlation length. These findings, together with the recent results on dynamic (inertial) 3D fracture~\cite{lubomirsky2023quenched}, highlight the essential roles played by quenched disorder in 3D fracture of amorphous materials. We further showed that steps emerge from the background surface roughness and elucidated their topology/geometry as they evolve.

Our findings are in agreement with a broad range of experiments and provide a unifying physical picture of step formation. We particularly highlight the experiments of~\cite{baumberger2008magic,ronsin2014crack}, which stressed the importance of the background fluctuations/roughness and suggested a close relation between finite mode I+III mixity and nominally mode-I step formation, and those of~\cite{steps2017,wang2022hidden}, which provided great insight into the topology/geometry of steps.

Our results are qualitatively consistent with the suggestion that step formation is strongly subcritical with respect to a linear helical instability~\cite{leblond2011theoretical,chen2015,pons2010helical}, made in an attempt to reconcile the large linear instability threshold ($(K_{_{\rm III}}/K_{_{\rm I}})_{\rm c}\!\simeq\!0.46$ for our Poisson's ratio~\cite{SI}) with experiments that indicate a very small, global threshold. Yet, we also stress the differences between the nonlinear, disorder-induced, localized step formation process we discussed and the extended, quasi-sinusoidal linear instability discussed in~\cite{leblond2011theoretical,chen2015,pons2010helical}. Future theoretical developments should strongly depart from perturbative approaches and incorporate realistic quenched disorder.

The mesoscopic mode I+III mixity we employed corresponds to coherent mode-mixity over a simulation length $L_z$, which may either represent a mode-mixity fluctuation in a nominally mode-I experiment or part of a macroscopic system under global mode-mixity. In the former context, future work should consider mode-mixity that varies in space, including of different signs, and study its effect on step formation and dynamics.

Finally, future work should address issues such as step drift~\cite{baumberger2008magic,steinhardt2023geometric,steps2017}, step-step interactions~\cite{tanaka1998discontinuous,baumberger2008magic,ronsin2014crack,steinhardt2023geometric,steps2017} and the interplay between steps and micro-branches~\cite{steps2017}, which have not been addressed here. Moreover, steps have been observed both in soft materials (e.g.,~\cite{tanaka1998discontinuous,baumberger2008magic,ronsin2014crack,steinhardt2023geometric,steps2017}), where an intrinsic nonlinear elastic lengthscale~\cite{bouchbinder.08a,goldman2012,bouchbinder.14,lubomirsky2018,chen2017,vasudevan2021oscillatory} is expected to play a role in the step formation process, and in hard materials (e.g.,~\cite{sommer1969formation,pham2014further}), where other lengthscales are expected to play a role. The effects of these intrinsic lengthscales, e.g., on the asymptotic step height, should also be explored in future work.

%\clearpage

\vspace{2cm}

{\em Acknowledgements}. E.B.~dedicates this work in the memory of his beloved grandfather, Zalman Bouchbinder (1926-2024), who passed away during the final stages of preparation of the manuscript. This work has been supported by the United States-Israel Binational Science Foundation (BSF, grant 2018603). E.B.~acknowledges support from the Ben May Center for Chemical Theory and Computation, and the Harold Perlman Family.

\clearpage

\onecolumngrid
%\vspace{1cm}
\begin{center}
              \textbf{\Large Supplemental materials}
\end{center}

%%%%%%%%%%%%%%%%%%%%%%%%%%%%%%%%%%%%%%%%%%%%%%%%%%%%%%%%%%%%%%%%%%%%%%%%%%%%%%%%%
%%%%%%%%%%%%%%%%%%%%%% these lines of code handle the concatenation %%%%%%%%%%%%%
%%%%%%%%%%%%%%%%%%%%%%%%%%%%%%%%%%%%%%%%%%%%%%%%%%%%%%%%%%%%%%%%%%%%%%%%%%%%%%%%%
\setcounter{equation}{0}
\setcounter{figure}{0}
\setcounter{section}{0}
\setcounter{subsection}{0}
\setcounter{table}{0}
\setcounter{page}{1}
\makeatletter
\renewcommand{\theequation}{S\arabic{equation}}
\renewcommand{\thefigure}{S\arabic{figure}}
\renewcommand{\thesection}{S-\Roman{section}}
\renewcommand{\thesubsection}{S-\Roman{subsection}}
\renewcommand*{\thepage}{S\arabic{page}}
%\renewcommand{\bibnumfmt}[1]{[S#1]}
%\renewcommand{\citenumfont}[1]{S#1}
%%%%%%%%%%%%%%%%%%%%%%%%%%%%%%%%%%%%%%%%%%%%%%%%%%%%%%%%%%%%%%%%%%%%%%%%%%%%%%%%%
%%%%%%%%%%%%%%%%%%%%%% these lines of code handle the concatenation %%%%%%%%%%%%%
%%%%%%%%%%%%%%%%%%%%%%%%%%%%%%%%%%%%%%%%%%%%%%%%%%%%%%%%%%%%%%%%%%%%%%%%%%%%%%%%%
\twocolumngrid

The goal of this document is to provide some technical details regarding the results presented in the manuscript and to offer some additional supporting data.\\

The 3D phase-field fracture framework is identical to the one detailed in~\cite{das2023dynamics,lubomirsky2023quenched}. Here, for completeness, we very briefly repeat the main elements of the formulation and highlight its main merits in the context of the present work. A general material is described in this framework by the following potential energy $U$, kinetic energy $T$ and dissipation function $D$~\cite{das2023dynamics,lubomirsky2023quenched,chen2017,lubomirsky2018,vasudevan2021oscillatory}
\begin{eqnarray}
\label{Eq:Lagrangian_U}
U&=&\int \left[\frac{1}{2}\kappa\left(\nabla\phi\right)^{2}+ g(\phi)\,e({\bm u}) + w(\phi)\,e_{\rm c}\right]dV \ ,\quad \\
\label{Eq:Lagrangian_T}
T&=&\int\!\frac{1}{2}f(\phi)\,\rho\left(\partial_t {\bm u}\right)^2 dV \ ,\quad \\
\label{Eq:dissipation}
D&=&\frac{1}{2\chi}\int \left(\partial_t \phi\right)^{2}dV \ ,\quad
\end{eqnarray}
in terms of a 3D time-dependent vectorial displacement field $\bm{u}(\bm{x},t)$ and a 3D time-dependent auxiliary scalar phase-field $0\!\le\!\phi(\bm{x},t)\!\le\!1$ ($\bm{x}\=(x,y,z)$ are Cartesian coordinates). Here, $dV$ is a volume differential and the integration extends over the entire system. The evolution of $\phi({\bm x},t)$ and ${\bm u}({\bm x},t)$ follows Lagrange's equations
\begin{eqnarray}
\frac{\partial}{\partial t}\left[\frac{\delta L}{\delta\left(\partial\psi/\partial t\right)}\right]-\frac{\delta L}{\delta\psi}
+\frac{\delta D}{\delta\left(\partial\psi/\partial t\right)}=0 \ .
\label{Eq:Lagrange_eqs}
\end{eqnarray}
Here $L\=T-U$ is the Lagrangian and $\psi\=(\phi,u_x,u_y,u_z)$, where ${\bm u}\=(u_x,u_y,u_z)$ are the components of the displacement vector field.

An intact/unbroken material state corresponds to $\phi\=1$, for which $g(\phi)\=f(1)\=1\!-\!w(1)\=1$. It describes a non-dissipative, elastic material response characterized by an energy density $e({\bm u})$. For the latter, we use the linear elastic energy density
\begin{equation}
e({\bm u})=\frac{1}{2}\lambda\,\text{tr}^2({\bm \varepsilon}) + \mu\,\text{tr}({\bm \varepsilon}) \ ,
\label{eq:LE}
\end{equation}
where ${\bm \varepsilon}\=\tfrac{1}{2}[{\bm \nabla}{\bm u}+({\bm \nabla}{\bm u})^{\rm T}]$ is the infinitesimal (linearized) strain tensor, and $\lambda$ and $\mu$ (shear modulus) are the Lam\'e coefficients. We set $\lambda\=2\mu$ in all of our calculations.

Dissipation, loss of load-bearing capacity and eventually fracture accompanied by the generation of traction-free surfaces are associated with a strain energy density threshold $e_{\rm c}$. When the latter is surpassed, $\phi$ decreases from unity and the degradation functions $g(\phi)$, $f(\phi)$ and $1\!-\!w(\phi)$ also decrease from unity towards zero, upon which fracture takes place. We adopt the so-called KKL choice of the degradation functions~\cite{karma2001phase,vasudevan2021oscillatory}, corresponding to $f(\phi)\=g(\phi)$ and $w(\phi)\=1-g(\phi)$, with $g(\phi)\=4\phi^3-3\phi^4$. The main merit of the phase-field framework is that it self-consistently selects both the near crack front dissipation and the front spatiotemporal evolution in 3D, without invoking any extraneous criteria (e.g., for crack path selection). In particular, tracking a small value iso-surface of the phase-field, conventionally $\phi(\bm{x},t)\=1/2$, allows to obtain the in silico real-time 3D spatiotemporal dynamics of cracks in a way that goes well beyond current experiments.

The resulting set of nonlinear partial differential field equations feature a dissipation lengthscale $\xi\=\sqrt{\kappa/2e_{\rm c}}$ near crack fronts and an associated dissipation timescale $\tau\=(2\chi e_{\rm c})^{-1}$. Upon expressing length in units of $\xi$, time in units of $\xi/c_{\rm s}$, energy density in units of $\mu$ and the mass density $\rho$ in units of $\mu/c_{\rm s}^2$ ($c_{\rm s}\=\sqrt{\mu/\rho}$ is the shear wave-speed), the dimensionless set of equations depends on just two dimensionless parameters: $e_{\rm c}/\mu$ and $\beta\=\tau\,c_{\rm s}/\xi$. The latter controls the $v$ dependence of the fracture energy, $\Gamma(v)$~\cite{chen2017,lubomirsky2018,vasudevan2021oscillatory}. In our calculations, we set $e_{\rm c}/\mu\=0.01$ and $\beta\=13.8$ (a larger value compared to other applications of the framework, see discussion below). We also set $L_y\=192\xi$ and $L_z\=318\xi$, where $L_x$ is essentially indefinitely large due to an employed treadmill procedure~\cite{vasudevan2021oscillatory} (the simulation box size in the $x$ direction is $192\xi$). The actual value of $L_x$ is limited by the simulation run time, which is typically very large (see below). The boundary conditions are specified below and the details of the numerical implementation are provided in~\cite{das2023dynamics,lubomirsky2023quenched}.

Quenched disorder is incorporated into the static fracture energy, exactly as described in~\cite{lubomirsky2023quenched}. It results in $\Gamma({\bm x};v\=0)/\Gamma_0$, which is a Gaussian quenched disorder field characterized by unity mean, standard deviation proportional to $\sigma$ (see~\cite{lubomirsky2023quenched} for a discussion of the proportionality prefactor) and spatial correlation length $R$. An example of the quenched disorder field is presented in~\cite{lubomirsky2023quenched}. Each calculation is performed on a single GPU (NVIDIA RTX A6000 or NVIDIA A40), either owned by the Bouchbinder group or available on one of the computer clusters at Weizmenn Institute of Science. A typical simulation time is $\sim\!5$ days.

As noted in the manuscript, there are only two differences in the present application of the phase-field framework compared to~\cite{lubomirsky2023quenched}. First, we add on top of the dominant mode-I (tensile) loading a small mode-III (anti-plane shear) loading component. Specifically, we set the displacement boundary conditions to be $u_y(x,y\!=\!\pm L_y/2,z,t)\=\Delta_{_{\rm I}}/2$ (mode-I) and $u_z(x,y\!=\!\pm L_y/2,z,t)\=\Delta_{_{\rm III}}/2$ (mode-III). Mode I+III mixity is commonly expressed as the ratio of the respective stress intensity factors~\cite{lawn}, $K_{_{\rm I}}$ and $K_{_{\rm III}}$, which quantify the intensity of the linear elastic $1/\sqrt{r}$ stress singularity associated with each symmetry mode. They are given by~\cite{lawn}
\begin{equation}
K_{_{\rm I}}\!=\!2\mu\,\Delta_{_{\rm I}}\sqrt{\frac{1+(\lambda/\mu)}{L_y}}, \quad\qquad K_{_{\rm III}}\!=\!\mu\,\Delta_{_{\rm III}}\sqrt{\frac{2}{L_y}} \ ,
\label{eq:SIFs}
\end{equation}
implying that
\begin{equation}
\frac{K_{_{\rm III}}}{K_{_{\rm I}}}\!=\!\frac{\Delta_{_{\rm III}}}{\Delta_{_{\rm I}}}\sqrt{\frac{1}{2\left[1+(\lambda/\mu)\right]}} \ .
\label{eq:mode_mixity}
\end{equation}
We consider small mode I+III mixity values, in the range $0\!\le\!K_{_{\rm III}}/K_{_{\rm I}}\!\le\!0.2$. Finally, the mode I+III mixity $K_{_{\rm III}}/K_{_{\rm I}}$ (which depends on the two loading displacements, $\Delta_{_{\rm I}}$ and $\Delta_{_{\rm III}}$) is varied under the constraint of a fixed a energy release rate (crack driving force) $G/\Gamma_0\=2.25$, where $G\=\left[(1-\nu)K_{_{\rm I}}^2+K_{_{\rm III}}^2\right]\!/2\mu$~\cite{freund,lawn}. Consequently, our loading conditions are controlled by a single parameter.

In the manuscript, we mention the linear stability threshold $(K_{_{\rm III}}/K_{_{\rm I}})_{\rm c}\!\simeq\!0.46$ for triggering an extended, quasi-sinusoidal instability in infinite systems~\cite{leblond2011theoretical,chen2015}. The latter corresponds to the expression $(K_{_{\rm III}}/K_{_{\rm I}})_{\rm c}(\nu)\=\sqrt{\frac{(1-\nu)(2-3\nu)}{3(2-\nu)-4\sqrt{2}(1-2\nu)}}$~\cite{leblond2011theoretical,chen2015}, which for our value of $\nu\=\lambda/[2(\lambda+\mu)]\=1/[2(1+(\mu/\lambda))]\=1/3$ (using $\lambda\=2\mu$, see above), gives $(K_{_{\rm III}}/K_{_{\rm I}})_{\rm c}(\nu\=1/3)\=0.4627\!\simeq\!0.46$.

\begin{figure}[ht!]
\center
\includegraphics[width=0.5\textwidth]{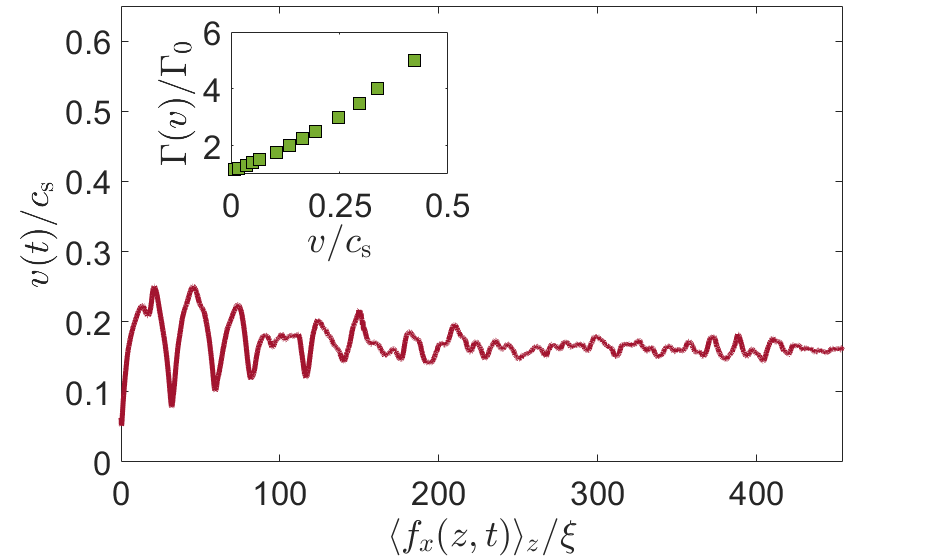}
%\vspace{-0.2cm}
\caption{$v(t)\!=\!\langle\partial_t f_x(z,t)\rangle_z$ (in units of $c_{\rm s}$) vs.~$\langle f_x(z,t)\rangle_z/\xi$ of the crack shown in Fig.~2a in the manuscript. (inset) $\Gamma(v)/\Gamma_0$ vs.~$v/c_{\rm s}$ for a homogeneous material ($\sigma\!=\!0$).}
\label{fig:figS1}
\end{figure}

The second difference compared to~\cite{lubomirsky2023quenched} is that we study slow cracks, i.e., we consider small crack propagation velocities in the quasi-static, non-inertial regime, because fracture surface steps were experimentally observed in this regime. An example of the crack velocity time evolution (corresponding to Fig.~2a in the manuscript) is presented in Fig.~\ref{fig:figS1}. Therein, we plot the reduced average crack front velocity $v(t)/c_{\rm s}$, with $v(t)\=\langle\partial_t f_x(z,t)\rangle_z$, against $\langle f_x(z,t)\rangle_z/\xi$ (where $\langle\,\cdot\,\rangle_z$ is an average over the thickness dimension $z$). It is observed that $v(t)\!\ll\!c_{\rm s}$ at all times (the asymptotic velocity is $\simeq\!0.16c_{\rm s}$). Quasi-static crack propagation is achieved both by selecting the crack driving force $G/\Gamma_0\=2.25$ and by using a rather large value of $\beta$, i.e., $\beta\=14$ (as stated above), such that $d\Gamma(v)/dv$ is large. The resulting $v$-dependent fracture energy $\Gamma(v)$ (for a homogeneous material) is plotted in the inset of Fig.~\ref{fig:figS1}.
%%%%%%%%%%%%%%%%%%%%%%%%%%%%%%%%%%%%%%%
%%%%%%%%%%%%%%%%%%%%%%%%%%%%%%%%%%%%%%%

As stated above, all of the results reported on in the manuscript were obtained with $L_y\=192\xi$ and $L_z\=318\xi$, i.e., $L_y/L_z\!\simeq\!0.6$. These are very large-scale computations performed on cutting-edge GPUs, which run for $\sim\!5$ days each. We are unable to further increase the system size using single-GPU calculations. Yet, in order to test for possible finite-size effects in our results, we performed calculations with $L_y/L_z\=2$ at fixed $\sigma\=0.5$ and $R/\xi\=10$, and variable $K_{_{\rm III}}/K_{_{\rm I}}$. We found that the onset of steps occurs at a mode I+III mixity level similar to the one obtained with $L_y/L_z\!\simeq\!0.6$ (reported on in Fig.~3a in the manuscript), indicating weak/negligble finite-size effects.


%\bibliography{unified3}

\begin{thebibliography}{56}%
\makeatletter
\providecommand \@ifxundefined [1]{%
 \@ifx{#1\undefined}
}%
\providecommand \@ifnum [1]{%
 \ifnum #1\expandafter \@firstoftwo
 \else \expandafter \@secondoftwo
 \fi
}%
\providecommand \@ifx [1]{%
 \ifx #1\expandafter \@firstoftwo
 \else \expandafter \@secondoftwo
 \fi
}%
\providecommand \natexlab [1]{#1}%
\providecommand \enquote  [1]{``#1''}%
\providecommand \bibnamefont  [1]{#1}%
\providecommand \bibfnamefont [1]{#1}%
\providecommand \citenamefont [1]{#1}%
\providecommand \href@noop [0]{\@secondoftwo}%
\providecommand \href [0]{\begingroup \@sanitize@url \@href}%
\providecommand \@href[1]{\@@startlink{#1}\@@href}%
\providecommand \@@href[1]{\endgroup#1\@@endlink}%
\providecommand \@sanitize@url [0]{\catcode `\\12\catcode `\$12\catcode
  `\&12\catcode `\#12\catcode `\^12\catcode `\_12\catcode `\%12\relax}%
\providecommand \@@startlink[1]{}%
\providecommand \@@endlink[0]{}%
\providecommand \url  [0]{\begingroup\@sanitize@url \@url }%
\providecommand \@url [1]{\endgroup\@href {#1}{\urlprefix }}%
\providecommand \urlprefix  [0]{URL }%
\providecommand \Eprint [0]{\href }%
\providecommand \doibase [0]{https://doi.org/}%
\providecommand \selectlanguage [0]{\@gobble}%
\providecommand \bibinfo  [0]{\@secondoftwo}%
\providecommand \bibfield  [0]{\@secondoftwo}%
\providecommand \translation [1]{[#1]}%
\providecommand \BibitemOpen [0]{}%
\providecommand \bibitemStop [0]{}%
\providecommand \bibitemNoStop [0]{.\EOS\space}%
\providecommand \EOS [0]{\spacefactor3000\relax}%
\providecommand \BibitemShut  [1]{\csname bibitem#1\endcsname}%
\let\auto@bib@innerbib\@empty
%</preamble>
\bibitem [{\citenamefont {Gent}\ and\ \citenamefont
  {Pulford}(1984)}]{gent1984micromechanics}%
  \BibitemOpen
  \bibfield  {author} {\bibinfo {author} {\bibfnamefont {A.}~\bibnamefont
  {Gent}}\ and\ \bibinfo {author} {\bibfnamefont {C.}~\bibnamefont {Pulford}},\
  }\bibfield  {title} {\bibinfo {title} {Micromechanics of fracture in
  elastomers},\ }\href@noop {} {\bibfield  {journal} {\bibinfo  {journal}
  {Journal of Materials Science}\ }\textbf {\bibinfo {volume} {19}},\ \bibinfo
  {pages} {3612} (\bibinfo {year} {1984})}\BibitemShut {NoStop}%
\bibitem [{\citenamefont {Tanaka}\ \emph {et~al.}(1998)\citenamefont {Tanaka},
  \citenamefont {Fukao}, \citenamefont {Miyamoto},\ and\ \citenamefont
  {Sekimoto}}]{tanaka1998discontinuous}%
  \BibitemOpen
  \bibfield  {author} {\bibinfo {author} {\bibfnamefont {Y.}~\bibnamefont
  {Tanaka}}, \bibinfo {author} {\bibfnamefont {K.}~\bibnamefont {Fukao}},
  \bibinfo {author} {\bibfnamefont {Y.}~\bibnamefont {Miyamoto}},\ and\
  \bibinfo {author} {\bibfnamefont {K.}~\bibnamefont {Sekimoto}},\ }\bibfield
  {title} {\bibinfo {title} {Discontinuous crack fronts of three-dimensional
  fractures},\ }\href@noop {} {\bibfield  {journal} {\bibinfo  {journal} {EPL
  (Europhysics Letters)}\ }\textbf {\bibinfo {volume} {43}},\ \bibinfo {pages}
  {664} (\bibinfo {year} {1998})}\BibitemShut {NoStop}%
\bibitem [{\citenamefont {Tanaka}\ \emph {et~al.}(2000)\citenamefont {Tanaka},
  \citenamefont {Fukao},\ and\ \citenamefont {Miyamoto}}]{tanaka2000fracture}%
  \BibitemOpen
  \bibfield  {author} {\bibinfo {author} {\bibfnamefont {Y.}~\bibnamefont
  {Tanaka}}, \bibinfo {author} {\bibfnamefont {K.}~\bibnamefont {Fukao}},\ and\
  \bibinfo {author} {\bibfnamefont {Y.}~\bibnamefont {Miyamoto}},\ }\bibfield
  {title} {\bibinfo {title} {Fracture energy of gels},\ }\href@noop {}
  {\bibfield  {journal} {\bibinfo  {journal} {The European Physical Journal E}\
  }\textbf {\bibinfo {volume} {3}},\ \bibinfo {pages} {395} (\bibinfo {year}
  {2000})}\BibitemShut {NoStop}%
\bibitem [{\citenamefont {Baumberger}\ \emph {et~al.}(2008)\citenamefont
  {Baumberger}, \citenamefont {Caroli}, \citenamefont {Martina},\ and\
  \citenamefont {Ronsin}}]{baumberger2008magic}%
  \BibitemOpen
  \bibfield  {author} {\bibinfo {author} {\bibfnamefont {T.}~\bibnamefont
  {Baumberger}}, \bibinfo {author} {\bibfnamefont {C.}~\bibnamefont {Caroli}},
  \bibinfo {author} {\bibfnamefont {D.}~\bibnamefont {Martina}},\ and\ \bibinfo
  {author} {\bibfnamefont {O.}~\bibnamefont {Ronsin}},\ }\bibfield  {title}
  {\bibinfo {title} {Magic angles and cross-hatching instability in hydrogel
  fracture},\ }\href@noop {} {\bibfield  {journal} {\bibinfo  {journal}
  {Physical Review Letters}\ }\textbf {\bibinfo {volume} {100}},\ \bibinfo
  {pages} {178303} (\bibinfo {year} {2008})}\BibitemShut {NoStop}%
\bibitem [{\citenamefont {Ronsin}\ \emph {et~al.}(2014)\citenamefont {Ronsin},
  \citenamefont {Caroli},\ and\ \citenamefont {Baumberger}}]{ronsin2014crack}%
  \BibitemOpen
  \bibfield  {author} {\bibinfo {author} {\bibfnamefont {O.}~\bibnamefont
  {Ronsin}}, \bibinfo {author} {\bibfnamefont {C.}~\bibnamefont {Caroli}},\
  and\ \bibinfo {author} {\bibfnamefont {T.}~\bibnamefont {Baumberger}},\
  }\bibfield  {title} {\bibinfo {title} {Crack front echelon instability in
  mixed mode fracture of a strongly nonlinear elastic solid},\ }\href@noop {}
  {\bibfield  {journal} {\bibinfo  {journal} {EPL (Europhysics Letters)}\
  }\textbf {\bibinfo {volume} {105}},\ \bibinfo {pages} {34001} (\bibinfo
  {year} {2014})}\BibitemShut {NoStop}%
\bibitem [{\citenamefont {Kolvin}\ \emph {et~al.}(2018)\citenamefont {Kolvin},
  \citenamefont {Cohen},\ and\ \citenamefont {Fineberg}}]{steps2017}%
  \BibitemOpen
  \bibfield  {author} {\bibinfo {author} {\bibfnamefont {I.}~\bibnamefont
  {Kolvin}}, \bibinfo {author} {\bibfnamefont {G.}~\bibnamefont {Cohen}},\ and\
  \bibinfo {author} {\bibfnamefont {J.}~\bibnamefont {Fineberg}},\ }\bibfield
  {title} {\bibinfo {title} {Topological defects govern crack front motion and
  facet formation on broken surfaces},\ }\href@noop {} {\bibfield  {journal}
  {\bibinfo  {journal} {Nature Materials}\ }\textbf {\bibinfo {volume} {17}},\
  \bibinfo {pages} {140} (\bibinfo {year} {2018})}\BibitemShut {NoStop}%
\bibitem [{\citenamefont {Steinhardt}\ and\ \citenamefont
  {Rubinstein}(2022)}]{steinhardt2022material}%
  \BibitemOpen
  \bibfield  {author} {\bibinfo {author} {\bibfnamefont {W.}~\bibnamefont
  {Steinhardt}}\ and\ \bibinfo {author} {\bibfnamefont {S.~M.}\ \bibnamefont
  {Rubinstein}},\ }\bibfield  {title} {\bibinfo {title} {How material
  heterogeneity creates rough fractures},\ }\href@noop {} {\bibfield  {journal}
  {\bibinfo  {journal} {Physical Review Letters}\ }\textbf {\bibinfo {volume}
  {129}},\ \bibinfo {pages} {128001} (\bibinfo {year} {2022})}\BibitemShut
  {NoStop}%
\bibitem [{\citenamefont {Wang}\ \emph {et~al.}(2022)\citenamefont {Wang},
  \citenamefont {Adda-Bedia}, \citenamefont {Kolinski},\ and\ \citenamefont
  {Fineberg}}]{wang2022hidden}%
  \BibitemOpen
  \bibfield  {author} {\bibinfo {author} {\bibfnamefont {M.}~\bibnamefont
  {Wang}}, \bibinfo {author} {\bibfnamefont {M.}~\bibnamefont {Adda-Bedia}},
  \bibinfo {author} {\bibfnamefont {J.~M.}\ \bibnamefont {Kolinski}},\ and\
  \bibinfo {author} {\bibfnamefont {J.}~\bibnamefont {Fineberg}},\ }\bibfield
  {title} {\bibinfo {title} {How hidden 3\uppercase{D} structure within crack
  fronts reveals energy balance},\ }\href@noop {} {\bibfield  {journal}
  {\bibinfo  {journal} {Journal of the Mechanics and Physics of Solids}\
  }\textbf {\bibinfo {volume} {161}},\ \bibinfo {pages} {104795} (\bibinfo
  {year} {2022})}\BibitemShut {NoStop}%
\bibitem [{\citenamefont {Steinhardt}\ and\ \citenamefont
  {Rubinstein}(2023)}]{steinhardt2023geometric}%
  \BibitemOpen
  \bibfield  {author} {\bibinfo {author} {\bibfnamefont {W.}~\bibnamefont
  {Steinhardt}}\ and\ \bibinfo {author} {\bibfnamefont {S.~M.}\ \bibnamefont
  {Rubinstein}},\ }\bibfield  {title} {\bibinfo {title} {Geometric rules for
  the annihilation dynamics of step lines on fracture fronts},\ }\href@noop {}
  {\bibfield  {journal} {\bibinfo  {journal} {Physical Review E}\ }\textbf
  {\bibinfo {volume} {107}},\ \bibinfo {pages} {055003} (\bibinfo {year}
  {2023})}\BibitemShut {NoStop}%
\bibitem [{\citenamefont {Wang}\ \emph {et~al.}(2023)\citenamefont {Wang},
  \citenamefont {Adda-Bedia},\ and\ \citenamefont
  {Fineberg}}]{wang2023dynamics}%
  \BibitemOpen
  \bibfield  {author} {\bibinfo {author} {\bibfnamefont {M.}~\bibnamefont
  {Wang}}, \bibinfo {author} {\bibfnamefont {M.}~\bibnamefont {Adda-Bedia}},\
  and\ \bibinfo {author} {\bibfnamefont {J.}~\bibnamefont {Fineberg}},\
  }\bibfield  {title} {\bibinfo {title} {Dynamics of three-dimensional stepped
  cracks, bistability, and their transition to simple cracks},\ }\href@noop {}
  {\bibfield  {journal} {\bibinfo  {journal} {Physical Review Research}\
  }\textbf {\bibinfo {volume} {5}},\ \bibinfo {pages} {L012001} (\bibinfo
  {year} {2023})}\BibitemShut {NoStop}%
\bibitem [{\citenamefont {Pham}\ and\ \citenamefont
  {Ravi-Chandar}(2014)}]{pham2014further}%
  \BibitemOpen
  \bibfield  {author} {\bibinfo {author} {\bibfnamefont {K.}~\bibnamefont
  {Pham}}\ and\ \bibinfo {author} {\bibfnamefont {K.}~\bibnamefont
  {Ravi-Chandar}},\ }\bibfield  {title} {\bibinfo {title} {Further examination
  of the criterion for crack initiation under mixed-mode \uppercase{I+III}
  loading},\ }\href@noop {} {\bibfield  {journal} {\bibinfo  {journal}
  {International Journal of Fracture}\ }\textbf {\bibinfo {volume} {189}},\
  \bibinfo {pages} {121} (\bibinfo {year} {2014})}\BibitemShut {NoStop}%
\bibitem [{\citenamefont {Ravi-Chandar}\ and\ \citenamefont
  {Knauss}(1984{\natexlab{a}})}]{ravi1984experimental_II}%
  \BibitemOpen
  \bibfield  {author} {\bibinfo {author} {\bibfnamefont {K.}~\bibnamefont
  {Ravi-Chandar}}\ and\ \bibinfo {author} {\bibfnamefont {W.}~\bibnamefont
  {Knauss}},\ }\bibfield  {title} {\bibinfo {title} {An experimental
  investigation into dynamic fracture: \uppercase{II. M}icrostructural
  aspects},\ }\href@noop {} {\bibfield  {journal} {\bibinfo  {journal}
  {International Journal of fracture}\ }\textbf {\bibinfo {volume} {26}},\
  \bibinfo {pages} {65} (\bibinfo {year} {1984}{\natexlab{a}})}\BibitemShut
  {NoStop}%
\bibitem [{\citenamefont {Ravi-Chandar}\ and\ \citenamefont
  {Knauss}(1984{\natexlab{b}})}]{ravi1984experimental_III}%
  \BibitemOpen
  \bibfield  {author} {\bibinfo {author} {\bibfnamefont {K.}~\bibnamefont
  {Ravi-Chandar}}\ and\ \bibinfo {author} {\bibfnamefont {W.~G.}\ \bibnamefont
  {Knauss}},\ }\bibfield  {title} {\bibinfo {title} {An experimental
  investigation into dynamic fracture: \uppercase{III. O}n steady-state crack
  propagation and crack branching},\ }\href@noop {} {\bibfield  {journal}
  {\bibinfo  {journal} {International Journal of fracture}\ }\textbf {\bibinfo
  {volume} {26}},\ \bibinfo {pages} {141} (\bibinfo {year}
  {1984}{\natexlab{b}})}\BibitemShut {NoStop}%
\bibitem [{\citenamefont {Scheibert}\ \emph {et~al.}(2010)\citenamefont
  {Scheibert}, \citenamefont {Guerra}, \citenamefont {C{\'e}lari{\'e}},
  \citenamefont {Dalmas},\ and\ \citenamefont {Bonamy}}]{scheibert2010brittle}%
  \BibitemOpen
  \bibfield  {author} {\bibinfo {author} {\bibfnamefont {J.}~\bibnamefont
  {Scheibert}}, \bibinfo {author} {\bibfnamefont {C.}~\bibnamefont {Guerra}},
  \bibinfo {author} {\bibfnamefont {F.}~\bibnamefont {C{\'e}lari{\'e}}},
  \bibinfo {author} {\bibfnamefont {D.}~\bibnamefont {Dalmas}},\ and\ \bibinfo
  {author} {\bibfnamefont {D.}~\bibnamefont {Bonamy}},\ }\bibfield  {title}
  {\bibinfo {title} {Brittle-quasibrittle transition in dynamic fracture: An
  energetic signature},\ }\href@noop {} {\bibfield  {journal} {\bibinfo
  {journal} {Physical Review Letters}\ }\textbf {\bibinfo {volume} {104}},\
  \bibinfo {pages} {045501} (\bibinfo {year} {2010})}\BibitemShut {NoStop}%
\bibitem [{\citenamefont {Guerra}\ \emph {et~al.}(2012)\citenamefont {Guerra},
  \citenamefont {Scheibert}, \citenamefont {Bonamy},\ and\ \citenamefont
  {Dalmas}}]{guerra2012understanding}%
  \BibitemOpen
  \bibfield  {author} {\bibinfo {author} {\bibfnamefont {C.}~\bibnamefont
  {Guerra}}, \bibinfo {author} {\bibfnamefont {J.}~\bibnamefont {Scheibert}},
  \bibinfo {author} {\bibfnamefont {D.}~\bibnamefont {Bonamy}},\ and\ \bibinfo
  {author} {\bibfnamefont {D.}~\bibnamefont {Dalmas}},\ }\bibfield  {title}
  {\bibinfo {title} {Understanding fast macroscale fracture from microcrack
  post mortem patterns},\ }\href@noop {} {\bibfield  {journal} {\bibinfo
  {journal} {Proceedings of the National Academy of Sciences}\ }\textbf
  {\bibinfo {volume} {109}},\ \bibinfo {pages} {390} (\bibinfo {year}
  {2012})}\BibitemShut {NoStop}%
\bibitem [{\citenamefont {Sharon}\ and\ \citenamefont
  {Fineberg}(1996)}]{sharon1996microbranching}%
  \BibitemOpen
  \bibfield  {author} {\bibinfo {author} {\bibfnamefont {E.}~\bibnamefont
  {Sharon}}\ and\ \bibinfo {author} {\bibfnamefont {J.}~\bibnamefont
  {Fineberg}},\ }\bibfield  {title} {\bibinfo {title} {Microbranching
  instability and the dynamic fracture of brittle materials},\ }\href@noop {}
  {\bibfield  {journal} {\bibinfo  {journal} {Physical Review B}\ }\textbf
  {\bibinfo {volume} {54}},\ \bibinfo {pages} {7128} (\bibinfo {year}
  {1996})}\BibitemShut {NoStop}%
\bibitem [{\citenamefont {Sharon}\ and\ \citenamefont
  {Fineberg}(1998)}]{sharon1998universal}%
  \BibitemOpen
  \bibfield  {author} {\bibinfo {author} {\bibfnamefont {E.}~\bibnamefont
  {Sharon}}\ and\ \bibinfo {author} {\bibfnamefont {J.}~\bibnamefont
  {Fineberg}},\ }\bibfield  {title} {\bibinfo {title} {Universal features of
  the microbranching instability in dynamic fracture},\ }\href@noop {}
  {\bibfield  {journal} {\bibinfo  {journal} {Philosophical Magazine B}\
  }\textbf {\bibinfo {volume} {78}},\ \bibinfo {pages} {243} (\bibinfo {year}
  {1998})}\BibitemShut {NoStop}%
\bibitem [{\citenamefont {Sharon}\ and\ \citenamefont
  {Fineberg}(1999)}]{sharon1999dynamics}%
  \BibitemOpen
  \bibfield  {author} {\bibinfo {author} {\bibfnamefont {E.}~\bibnamefont
  {Sharon}}\ and\ \bibinfo {author} {\bibfnamefont {J.}~\bibnamefont
  {Fineberg}},\ }\bibfield  {title} {\bibinfo {title} {The dynamics of fast
  fracture},\ }\href@noop {} {\bibfield  {journal} {\bibinfo  {journal}
  {Advanced Engineering Materials}\ }\textbf {\bibinfo {volume} {1}},\ \bibinfo
  {pages} {119} (\bibinfo {year} {1999})}\BibitemShut {NoStop}%
\bibitem [{\citenamefont {Livne}\ \emph {et~al.}(2005)\citenamefont {Livne},
  \citenamefont {Cohen},\ and\ \citenamefont
  {Fineberg}}]{livne2005universality}%
  \BibitemOpen
  \bibfield  {author} {\bibinfo {author} {\bibfnamefont {A.}~\bibnamefont
  {Livne}}, \bibinfo {author} {\bibfnamefont {G.}~\bibnamefont {Cohen}},\ and\
  \bibinfo {author} {\bibfnamefont {J.}~\bibnamefont {Fineberg}},\ }\bibfield
  {title} {\bibinfo {title} {Universality and hysteretic dynamics in rapid
  fracture},\ }\href@noop {} {\bibfield  {journal} {\bibinfo  {journal}
  {Physical Review Letters}\ }\textbf {\bibinfo {volume} {94}},\ \bibinfo
  {pages} {224301} (\bibinfo {year} {2005})}\BibitemShut {NoStop}%
\bibitem [{\citenamefont {Lawn}(1993)}]{lawn}%
  \BibitemOpen
  \bibfield  {author} {\bibinfo {author} {\bibfnamefont {B.}~\bibnamefont
  {Lawn}},\ }\href@noop {} {\emph {\bibinfo {title} {Fracture of Brittle
  Solids}}}\ (\bibinfo  {publisher} {Cambridge University Press},\ \bibinfo
  {year} {1993})\BibitemShut {NoStop}%
\bibitem [{\citenamefont {Johnson}\ and\ \citenamefont
  {Holloway}(1968)}]{johnson1968microstructure}%
  \BibitemOpen
  \bibfield  {author} {\bibinfo {author} {\bibfnamefont {J.}~\bibnamefont
  {Johnson}}\ and\ \bibinfo {author} {\bibfnamefont {D.}~\bibnamefont
  {Holloway}},\ }\bibfield  {title} {\bibinfo {title} {Microstructure of the
  mist zone on glass fracture surfaces},\ }\href@noop {} {\bibfield  {journal}
  {\bibinfo  {journal} {The Philosophical Magazine: A Journal of Theoretical
  Experimental and Applied Physics}\ }\textbf {\bibinfo {volume} {17}},\
  \bibinfo {pages} {899} (\bibinfo {year} {1968})}\BibitemShut {NoStop}%
\bibitem [{\citenamefont {Rabinovitch}\ \emph {et~al.}(2000)\citenamefont
  {Rabinovitch}, \citenamefont {Belizovsky},\ and\ \citenamefont
  {Bahat}}]{rabinovitch2000origin}%
  \BibitemOpen
  \bibfield  {author} {\bibinfo {author} {\bibfnamefont {A.}~\bibnamefont
  {Rabinovitch}}, \bibinfo {author} {\bibfnamefont {G.}~\bibnamefont
  {Belizovsky}},\ and\ \bibinfo {author} {\bibfnamefont {D.}~\bibnamefont
  {Bahat}},\ }\bibfield  {title} {\bibinfo {title} {Origin of mist and hackle
  patterns in brittle fracture},\ }\href@noop {} {\bibfield  {journal}
  {\bibinfo  {journal} {Physical Review B}\ }\textbf {\bibinfo {volume} {61}},\
  \bibinfo {pages} {14968} (\bibinfo {year} {2000})}\BibitemShut {NoStop}%
\bibitem [{\citenamefont {Jiao}\ \emph {et~al.}(2015)\citenamefont {Jiao},
  \citenamefont {Qu},\ and\ \citenamefont {Zhang}}]{jiao2015macroscopic}%
  \BibitemOpen
  \bibfield  {author} {\bibinfo {author} {\bibfnamefont {D.}~\bibnamefont
  {Jiao}}, \bibinfo {author} {\bibfnamefont {R.~T.}\ \bibnamefont {Qu}},\ and\
  \bibinfo {author} {\bibfnamefont {Z.~F.}\ \bibnamefont {Zhang}},\ }\bibfield
  {title} {\bibinfo {title} {Macroscopic bifurcation and fracture mechanism of
  polymethyl methacrylate},\ }\href@noop {} {\bibfield  {journal} {\bibinfo
  {journal} {Advanced Engineering Materials}\ }\textbf {\bibinfo {volume}
  {17}},\ \bibinfo {pages} {1454} (\bibinfo {year} {2015})}\BibitemShut
  {NoStop}%
\bibitem [{\citenamefont {Lubomirsky}\ and\ \citenamefont
  {Bouchbinder}(2023)}]{lubomirsky2023quenched}%
  \BibitemOpen
  \bibfield  {author} {\bibinfo {author} {\bibfnamefont {Y.}~\bibnamefont
  {Lubomirsky}}\ and\ \bibinfo {author} {\bibfnamefont {E.}~\bibnamefont
  {Bouchbinder}},\ }\bibfield  {title} {\bibinfo {title} {Quenched disorder and
  instability control dynamic fracture in three dimensions},\ }\href@noop {}
  {\bibfield  {journal} {\bibinfo  {journal} {arXiv preprint arXiv:2311.11692}\
  } (\bibinfo {year} {2023})}\BibitemShut {NoStop}%
\bibitem [{\citenamefont {Ball}\ and\ \citenamefont
  {Larralde}(1995)}]{ball1995three}%
  \BibitemOpen
  \bibfield  {author} {\bibinfo {author} {\bibfnamefont {R.}~\bibnamefont
  {Ball}}\ and\ \bibinfo {author} {\bibfnamefont {H.}~\bibnamefont
  {Larralde}},\ }\bibfield  {title} {\bibinfo {title} {Three-dimensional
  stability analysis of planar straight cracks propagating quasistatically
  under type i loading},\ }\href@noop {} {\bibfield  {journal} {\bibinfo
  {journal} {International Journal of Fracture}\ }\textbf {\bibinfo {volume}
  {71}},\ \bibinfo {pages} {365} (\bibinfo {year} {1995})}\BibitemShut
  {NoStop}%
\bibitem [{\citenamefont {Movchan}\ \emph {et~al.}(1998)\citenamefont
  {Movchan}, \citenamefont {Gao},\ and\ \citenamefont
  {Willis}}]{movchan1998perturbations}%
  \BibitemOpen
  \bibfield  {author} {\bibinfo {author} {\bibfnamefont {A.}~\bibnamefont
  {Movchan}}, \bibinfo {author} {\bibfnamefont {H.}~\bibnamefont {Gao}},\ and\
  \bibinfo {author} {\bibfnamefont {J.}~\bibnamefont {Willis}},\ }\bibfield
  {title} {\bibinfo {title} {On perturbations of plane cracks},\ }\href@noop {}
  {\bibfield  {journal} {\bibinfo  {journal} {International Journal of Solids
  and Structures}\ }\textbf {\bibinfo {volume} {35}},\ \bibinfo {pages} {3419}
  (\bibinfo {year} {1998})}\BibitemShut {NoStop}%
\bibitem [{\citenamefont {Leblond}\ \emph {et~al.}(2011)\citenamefont
  {Leblond}, \citenamefont {Karma},\ and\ \citenamefont
  {Lazarus}}]{leblond2011theoretical}%
  \BibitemOpen
  \bibfield  {author} {\bibinfo {author} {\bibfnamefont {J.}~\bibnamefont
  {Leblond}}, \bibinfo {author} {\bibfnamefont {A.}~\bibnamefont {Karma}},\
  and\ \bibinfo {author} {\bibfnamefont {V.}~\bibnamefont {Lazarus}},\
  }\bibfield  {title} {\bibinfo {title} {Theoretical analysis of crack front
  instability in mode \uppercase{I+III}},\ }\href@noop {} {\bibfield  {journal}
  {\bibinfo  {journal} {Journal of the Mechanics and Physics of Solids}\
  }\textbf {\bibinfo {volume} {59}},\ \bibinfo {pages} {1872} (\bibinfo {year}
  {2011})}\BibitemShut {NoStop}%
\bibitem [{\citenamefont {Leblond}\ and\ \citenamefont
  {Ponson}(2016)}]{leblond2016out}%
  \BibitemOpen
  \bibfield  {author} {\bibinfo {author} {\bibfnamefont {J.-B.}\ \bibnamefont
  {Leblond}}\ and\ \bibinfo {author} {\bibfnamefont {L.}~\bibnamefont
  {Ponson}},\ }\bibfield  {title} {\bibinfo {title} {Out-of-plane deviation of
  a mode \uppercase{I+III} crack encountering a tougher obstacle},\ }\href@noop
  {} {\bibfield  {journal} {\bibinfo  {journal} {Comptes Rendus.
  M{\'e}canique}\ }\textbf {\bibinfo {volume} {344}},\ \bibinfo {pages} {521}
  (\bibinfo {year} {2016})}\BibitemShut {NoStop}%
\bibitem [{\citenamefont {Lebihain}\ \emph {et~al.}(2020)\citenamefont
  {Lebihain}, \citenamefont {Leblond},\ and\ \citenamefont
  {Ponson}}]{lebihain2020effective}%
  \BibitemOpen
  \bibfield  {author} {\bibinfo {author} {\bibfnamefont {M.}~\bibnamefont
  {Lebihain}}, \bibinfo {author} {\bibfnamefont {J.-B.}\ \bibnamefont
  {Leblond}},\ and\ \bibinfo {author} {\bibfnamefont {L.}~\bibnamefont
  {Ponson}},\ }\bibfield  {title} {\bibinfo {title} {Effective toughness of
  periodic heterogeneous materials: the effect of out-of-plane excursions of
  cracks},\ }\href@noop {} {\bibfield  {journal} {\bibinfo  {journal} {Journal
  of the Mechanics and Physics of Solids}\ }\textbf {\bibinfo {volume} {137}},\
  \bibinfo {pages} {103876} (\bibinfo {year} {2020})}\BibitemShut {NoStop}%
\bibitem [{\citenamefont {Chen}\ \emph {et~al.}(2015)\citenamefont {Chen},
  \citenamefont {Cambonie}, \citenamefont {Lazarus}, \citenamefont {Nicoli},
  \citenamefont {Pons},\ and\ \citenamefont {Karma}}]{chen2015}%
  \BibitemOpen
  \bibfield  {author} {\bibinfo {author} {\bibfnamefont {C.-H.}\ \bibnamefont
  {Chen}}, \bibinfo {author} {\bibfnamefont {T.}~\bibnamefont {Cambonie}},
  \bibinfo {author} {\bibfnamefont {V.}~\bibnamefont {Lazarus}}, \bibinfo
  {author} {\bibfnamefont {M.}~\bibnamefont {Nicoli}}, \bibinfo {author}
  {\bibfnamefont {A.~J.}\ \bibnamefont {Pons}},\ and\ \bibinfo {author}
  {\bibfnamefont {A.}~\bibnamefont {Karma}},\ }\bibfield  {title} {\bibinfo
  {title} {Crack front segmentation and facet coarsening in mixed-mode
  fracture},\ }\href@noop {} {\bibfield  {journal} {\bibinfo  {journal}
  {Physical Review Letters}\ }\textbf {\bibinfo {volume} {115}},\ \bibinfo
  {pages} {265503} (\bibinfo {year} {2015})}\BibitemShut {NoStop}%
\bibitem [{\citenamefont {Pollard}\ \emph {et~al.}(1982)\citenamefont
  {Pollard}, \citenamefont {Segall},\ and\ \citenamefont
  {Delaney}}]{pollard1982formation}%
  \BibitemOpen
  \bibfield  {author} {\bibinfo {author} {\bibfnamefont {D.~D.}\ \bibnamefont
  {Pollard}}, \bibinfo {author} {\bibfnamefont {P.}~\bibnamefont {Segall}},\
  and\ \bibinfo {author} {\bibfnamefont {P.~T.}\ \bibnamefont {Delaney}},\
  }\bibfield  {title} {\bibinfo {title} {Formation and interpretation of
  dilatant echelon cracks},\ }\href@noop {} {\bibfield  {journal} {\bibinfo
  {journal} {Geological Society of America Bulletin}\ }\textbf {\bibinfo
  {volume} {93}},\ \bibinfo {pages} {1291} (\bibinfo {year}
  {1982})}\BibitemShut {NoStop}%
\bibitem [{\citenamefont {Lazarus}\ \emph {et~al.}(2008)\citenamefont
  {Lazarus}, \citenamefont {Buchholz}, \citenamefont {Fulland},\ and\
  \citenamefont {Wiebesiek}}]{lazarus2008comparison}%
  \BibitemOpen
  \bibfield  {author} {\bibinfo {author} {\bibfnamefont {V.}~\bibnamefont
  {Lazarus}}, \bibinfo {author} {\bibfnamefont {F.-G.}\ \bibnamefont
  {Buchholz}}, \bibinfo {author} {\bibfnamefont {M.}~\bibnamefont {Fulland}},\
  and\ \bibinfo {author} {\bibfnamefont {J.}~\bibnamefont {Wiebesiek}},\
  }\bibfield  {title} {\bibinfo {title} {Comparison of predictions by mode
  \uppercase{II} or mode \uppercase{III} criteria on crack front twisting in
  three or four point bending experiments},\ }\href@noop {} {\bibfield
  {journal} {\bibinfo  {journal} {International Journal of Fracture}\ }\textbf
  {\bibinfo {volume} {153}},\ \bibinfo {pages} {141} (\bibinfo {year}
  {2008})}\BibitemShut {NoStop}%
\bibitem [{\citenamefont {Pons}\ and\ \citenamefont
  {Karma}(2010)}]{pons2010helical}%
  \BibitemOpen
  \bibfield  {author} {\bibinfo {author} {\bibfnamefont {A.}~\bibnamefont
  {Pons}}\ and\ \bibinfo {author} {\bibfnamefont {A.}~\bibnamefont {Karma}},\
  }\bibfield  {title} {\bibinfo {title} {Helical crack-front instability in
  mixed-mode fracture},\ }\href@noop {} {\bibfield  {journal} {\bibinfo
  {journal} {Nature}\ }\textbf {\bibinfo {volume} {464}},\ \bibinfo {pages}
  {85} (\bibinfo {year} {2010})}\BibitemShut {NoStop}%
\bibitem [{\citenamefont {Goldstein}\ and\ \citenamefont
  {Osipenko}(2012)}]{goldstein2012fracture}%
  \BibitemOpen
  \bibfield  {author} {\bibinfo {author} {\bibfnamefont {R.}~\bibnamefont
  {Goldstein}}\ and\ \bibinfo {author} {\bibfnamefont {N.}~\bibnamefont
  {Osipenko}},\ }\bibfield  {title} {\bibinfo {title} {Fracture structure near
  a longitudinal shear macrorupture},\ }\href@noop {} {\bibfield  {journal}
  {\bibinfo  {journal} {Mechanics of Solids}\ }\textbf {\bibinfo {volume}
  {47}},\ \bibinfo {pages} {505} (\bibinfo {year} {2012})}\BibitemShut
  {NoStop}%
\bibitem [{\citenamefont {Leblond}\ \emph {et~al.}(2015)\citenamefont
  {Leblond}, \citenamefont {Lazarus},\ and\ \citenamefont
  {Karma}}]{leblond2015multiscale}%
  \BibitemOpen
  \bibfield  {author} {\bibinfo {author} {\bibfnamefont {J.-B.}\ \bibnamefont
  {Leblond}}, \bibinfo {author} {\bibfnamefont {V.}~\bibnamefont {Lazarus}},\
  and\ \bibinfo {author} {\bibfnamefont {A.}~\bibnamefont {Karma}},\ }\bibfield
   {title} {\bibinfo {title} {Multiscale cohesive zone model for propagation of
  segmented crack fronts in mode \uppercase{I+III} fracture},\ }\href@noop {}
  {\bibfield  {journal} {\bibinfo  {journal} {International Journal of
  Fracture}\ }\textbf {\bibinfo {volume} {191}},\ \bibinfo {pages} {167}
  (\bibinfo {year} {2015})}\BibitemShut {NoStop}%
\bibitem [{\citenamefont {Leblond}\ \emph {et~al.}(2019)\citenamefont
  {Leblond}, \citenamefont {Karma}, \citenamefont {Ponson},\ and\ \citenamefont
  {Vasudevan}}]{leblond2019configurational}%
  \BibitemOpen
  \bibfield  {author} {\bibinfo {author} {\bibfnamefont {J.-B.}\ \bibnamefont
  {Leblond}}, \bibinfo {author} {\bibfnamefont {A.}~\bibnamefont {Karma}},
  \bibinfo {author} {\bibfnamefont {L.}~\bibnamefont {Ponson}},\ and\ \bibinfo
  {author} {\bibfnamefont {A.}~\bibnamefont {Vasudevan}},\ }\bibfield  {title}
  {\bibinfo {title} {Configurational stability of a crack propagating in a
  material with mode-dependent fracture energy --- part \uppercase{I}:
  Mixed-mode \uppercase{I+III}},\ }\href@noop {} {\bibfield  {journal}
  {\bibinfo  {journal} {Journal of the Mechanics and Physics of Solids}\
  }\textbf {\bibinfo {volume} {126}},\ \bibinfo {pages} {187} (\bibinfo {year}
  {2019})}\BibitemShut {NoStop}%
\bibitem [{\citenamefont {Vasudevan}\ \emph {et~al.}(2020)\citenamefont
  {Vasudevan}, \citenamefont {Ponson}, \citenamefont {Karma},\ and\
  \citenamefont {Leblond}}]{vasudevan2020configurational}%
  \BibitemOpen
  \bibfield  {author} {\bibinfo {author} {\bibfnamefont {A.}~\bibnamefont
  {Vasudevan}}, \bibinfo {author} {\bibfnamefont {L.}~\bibnamefont {Ponson}},
  \bibinfo {author} {\bibfnamefont {A.}~\bibnamefont {Karma}},\ and\ \bibinfo
  {author} {\bibfnamefont {J.-B.}\ \bibnamefont {Leblond}},\ }\bibfield
  {title} {\bibinfo {title} {Configurational stability of a crack propagating
  in a material with mode-dependent fracture energy --- part \uppercase{II}:
  Drift of fracture facets in mixed-mode \uppercase{I+II+III}},\ }\href@noop {}
  {\bibfield  {journal} {\bibinfo  {journal} {Journal of the Mechanics and
  Physics of Solids}\ }\textbf {\bibinfo {volume} {137}},\ \bibinfo {pages}
  {103894} (\bibinfo {year} {2020})}\BibitemShut {NoStop}%
\bibitem [{\citenamefont {Lin}\ \emph {et~al.}(2010)\citenamefont {Lin},
  \citenamefont {Mear},\ and\ \citenamefont {Ravi-Chandar}}]{lin2010criterion}%
  \BibitemOpen
  \bibfield  {author} {\bibinfo {author} {\bibfnamefont {B.}~\bibnamefont
  {Lin}}, \bibinfo {author} {\bibfnamefont {M.}~\bibnamefont {Mear}},\ and\
  \bibinfo {author} {\bibfnamefont {K.}~\bibnamefont {Ravi-Chandar}},\
  }\bibfield  {title} {\bibinfo {title} {Criterion for initiation of cracks
  under mixed-mode \uppercase{I+III} loading},\ }\href@noop {} {\bibfield
  {journal} {\bibinfo  {journal} {International journal of fracture}\ }\textbf
  {\bibinfo {volume} {165}},\ \bibinfo {pages} {175} (\bibinfo {year}
  {2010})}\BibitemShut {NoStop}%
\bibitem [{\citenamefont {Pham}\ and\ \citenamefont
  {Ravi-Chandar}(2016)}]{pham2016growth}%
  \BibitemOpen
  \bibfield  {author} {\bibinfo {author} {\bibfnamefont {K.}~\bibnamefont
  {Pham}}\ and\ \bibinfo {author} {\bibfnamefont {K.}~\bibnamefont
  {Ravi-Chandar}},\ }\bibfield  {title} {\bibinfo {title} {On the growth of
  cracks under mixed-mode \uppercase{I+III} loading},\ }\href@noop {}
  {\bibfield  {journal} {\bibinfo  {journal} {International Journal of
  Fracture}\ }\textbf {\bibinfo {volume} {199}},\ \bibinfo {pages} {105}
  (\bibinfo {year} {2016})}\BibitemShut {NoStop}%
\bibitem [{\citenamefont {Pham}\ and\ \citenamefont
  {Ravi-Chandar}(2017)}]{pham2017formation}%
  \BibitemOpen
  \bibfield  {author} {\bibinfo {author} {\bibfnamefont {K.}~\bibnamefont
  {Pham}}\ and\ \bibinfo {author} {\bibfnamefont {K.}~\bibnamefont
  {Ravi-Chandar}},\ }\bibfield  {title} {\bibinfo {title} {The formation and
  growth of echelon cracks in brittle materials},\ }\href@noop {} {\bibfield
  {journal} {\bibinfo  {journal} {International Journal of Fracture}\ }\textbf
  {\bibinfo {volume} {206}},\ \bibinfo {pages} {229} (\bibinfo {year}
  {2017})}\BibitemShut {NoStop}%
\bibitem [{\citenamefont {Sommer}(1969)}]{sommer1969formation}%
  \BibitemOpen
  \bibfield  {author} {\bibinfo {author} {\bibfnamefont {E.}~\bibnamefont
  {Sommer}},\ }\bibfield  {title} {\bibinfo {title} {Formation of fracture
  ‘lances’ in glass},\ }\href@noop {} {\bibfield  {journal} {\bibinfo
  {journal} {Engineering Fracture Mechanics}\ }\textbf {\bibinfo {volume}
  {1}},\ \bibinfo {pages} {539} (\bibinfo {year} {1969})}\BibitemShut {NoStop}%
\bibitem [{\citenamefont {Lubomirsky}\ \emph {et~al.}(2018)\citenamefont
  {Lubomirsky}, \citenamefont {Chen}, \citenamefont {Karma},\ and\
  \citenamefont {Bouchbinder}}]{lubomirsky2018}%
  \BibitemOpen
  \bibfield  {author} {\bibinfo {author} {\bibfnamefont {Y.}~\bibnamefont
  {Lubomirsky}}, \bibinfo {author} {\bibfnamefont {C.-H.}\ \bibnamefont
  {Chen}}, \bibinfo {author} {\bibfnamefont {A.}~\bibnamefont {Karma}},\ and\
  \bibinfo {author} {\bibfnamefont {E.}~\bibnamefont {Bouchbinder}},\
  }\bibfield  {title} {\bibinfo {title} {Universality and stability phase
  diagram of two-dimensional brittle fracture},\ }\href@noop {} {\bibfield
  {journal} {\bibinfo  {journal} {Physical Review Letters}\ }\textbf {\bibinfo
  {volume} {121}},\ \bibinfo {pages} {134301} (\bibinfo {year}
  {2018})}\BibitemShut {NoStop}%
\bibitem [{\citenamefont {Chen}\ \emph {et~al.}(2017)\citenamefont {Chen},
  \citenamefont {Bouchbinder},\ and\ \citenamefont {Karma}}]{chen2017}%
  \BibitemOpen
  \bibfield  {author} {\bibinfo {author} {\bibfnamefont {C.-H.}\ \bibnamefont
  {Chen}}, \bibinfo {author} {\bibfnamefont {E.}~\bibnamefont {Bouchbinder}},\
  and\ \bibinfo {author} {\bibfnamefont {A.}~\bibnamefont {Karma}},\ }\bibfield
   {title} {\bibinfo {title} {Instability in dynamic fracture and the failure
  of the classical theory of cracks},\ }\href@noop {} {\bibfield  {journal}
  {\bibinfo  {journal} {Nature Physics}\ }\textbf {\bibinfo {volume} {13}},\
  \bibinfo {pages} {1186} (\bibinfo {year} {2017})}\BibitemShut {NoStop}%
\bibitem [{\citenamefont {Vasudevan}\ \emph {et~al.}(2021)\citenamefont
  {Vasudevan}, \citenamefont {Lubomirsky}, \citenamefont {Chen}, \citenamefont
  {Bouchbinder},\ and\ \citenamefont {Karma}}]{vasudevan2021oscillatory}%
  \BibitemOpen
  \bibfield  {author} {\bibinfo {author} {\bibfnamefont {A.}~\bibnamefont
  {Vasudevan}}, \bibinfo {author} {\bibfnamefont {Y.}~\bibnamefont
  {Lubomirsky}}, \bibinfo {author} {\bibfnamefont {C.-H.}\ \bibnamefont
  {Chen}}, \bibinfo {author} {\bibfnamefont {E.}~\bibnamefont {Bouchbinder}},\
  and\ \bibinfo {author} {\bibfnamefont {A.}~\bibnamefont {Karma}},\ }\bibfield
   {title} {\bibinfo {title} {Oscillatory and tip-splitting instabilities in
  \uppercase{2D} dynamic fracture: The roles of intrinsic material length and
  time scales},\ }\href@noop {} {\bibfield  {journal} {\bibinfo  {journal}
  {Journal of the Mechanics and Physics of Solids}\ }\textbf {\bibinfo {volume}
  {151}},\ \bibinfo {pages} {104372} (\bibinfo {year} {2021})}\BibitemShut
  {NoStop}%
\bibitem [{\citenamefont {Karma}\ \emph {et~al.}(2001)\citenamefont {Karma},
  \citenamefont {Kessler},\ and\ \citenamefont {Levine}}]{karma2001phase}%
  \BibitemOpen
  \bibfield  {author} {\bibinfo {author} {\bibfnamefont {A.}~\bibnamefont
  {Karma}}, \bibinfo {author} {\bibfnamefont {D.}~\bibnamefont {Kessler}},\
  and\ \bibinfo {author} {\bibfnamefont {H.}~\bibnamefont {Levine}},\
  }\bibfield  {title} {\bibinfo {title} {Phase-field model of mode
  \uppercase{III} dynamic fracture},\ }\href@noop {} {\bibfield  {journal}
  {\bibinfo  {journal} {Physical Review Letters}\ }\textbf {\bibinfo {volume}
  {87}},\ \bibinfo {pages} {45501} (\bibinfo {year} {2001})}\BibitemShut
  {NoStop}%
\bibitem [{\citenamefont {Karma}\ and\ \citenamefont
  {Lobkovsky}(2004)}]{Karma2004}%
  \BibitemOpen
  \bibfield  {author} {\bibinfo {author} {\bibfnamefont {A.}~\bibnamefont
  {Karma}}\ and\ \bibinfo {author} {\bibfnamefont {A.~E.}\ \bibnamefont
  {Lobkovsky}},\ }\bibfield  {title} {\bibinfo {title} {Unsteady crack motion
  and branching in a phase-field model of brittle fracture},\ }\href@noop {}
  {\bibfield  {journal} {\bibinfo  {journal} {Physical Review Letters}\
  }\textbf {\bibinfo {volume} {92}},\ \bibinfo {pages} {245510} (\bibinfo
  {year} {2004})}\BibitemShut {NoStop}%
\bibitem [{\citenamefont {Hakim}\ and\ \citenamefont {Karma}(2009)}]{Hakim.09}%
  \BibitemOpen
  \bibfield  {author} {\bibinfo {author} {\bibfnamefont {V.}~\bibnamefont
  {Hakim}}\ and\ \bibinfo {author} {\bibfnamefont {A.}~\bibnamefont {Karma}},\
  }\bibfield  {title} {\bibinfo {title} {Laws of crack motion and phase-field
  models of fracture},\ }\href@noop {} {\bibfield  {journal} {\bibinfo
  {journal} {Journal of the Mechanics and Physics of Solids}\ }\textbf
  {\bibinfo {volume} {57}},\ \bibinfo {pages} {342} (\bibinfo {year}
  {2009})}\BibitemShut {NoStop}%
\bibitem [{\citenamefont {Das}\ \emph {et~al.}(2023)\citenamefont {Das},
  \citenamefont {Lubomirsky},\ and\ \citenamefont
  {Bouchbinder}}]{das2023dynamics}%
  \BibitemOpen
  \bibfield  {author} {\bibinfo {author} {\bibfnamefont {S.}~\bibnamefont
  {Das}}, \bibinfo {author} {\bibfnamefont {Y.}~\bibnamefont {Lubomirsky}},\
  and\ \bibinfo {author} {\bibfnamefont {E.}~\bibnamefont {Bouchbinder}},\
  }\bibfield  {title} {\bibinfo {title} {Dynamics of crack front waves in
  three-dimensional material failure},\ }\href@noop {} {\bibfield  {journal}
  {\bibinfo  {journal} {Physical Review E}\ }\textbf {\bibinfo {volume}
  {108}},\ \bibinfo {pages} {L043002} (\bibinfo {year} {2023})}\BibitemShut
  {NoStop}%
\bibitem [{\citenamefont {Bleyer}\ and\ \citenamefont
  {Molinari}(2017)}]{bleyer2017microbranching}%
  \BibitemOpen
  \bibfield  {author} {\bibinfo {author} {\bibfnamefont {J.}~\bibnamefont
  {Bleyer}}\ and\ \bibinfo {author} {\bibfnamefont {J.-F.}\ \bibnamefont
  {Molinari}},\ }\bibfield  {title} {\bibinfo {title} {Microbranching
  instability in phase-field modelling of dynamic brittle fracture},\
  }\href@noop {} {\bibfield  {journal} {\bibinfo  {journal} {Applied Physics
  Letters}\ }\textbf {\bibinfo {volume} {110}},\ \bibinfo {pages} {151903}
  (\bibinfo {year} {2017})}\BibitemShut {NoStop}%
\bibitem [{\citenamefont {Henry}\ and\ \citenamefont
  {Adda-Bedia}(2013)}]{henry2013fractographic}%
  \BibitemOpen
  \bibfield  {author} {\bibinfo {author} {\bibfnamefont {H.}~\bibnamefont
  {Henry}}\ and\ \bibinfo {author} {\bibfnamefont {M.}~\bibnamefont
  {Adda-Bedia}},\ }\bibfield  {title} {\bibinfo {title} {Fractographic aspects
  of crack branching instability using a phase-field model},\ }\href@noop {}
  {\bibfield  {journal} {\bibinfo  {journal} {Physical Review E}\ }\textbf
  {\bibinfo {volume} {88}},\ \bibinfo {pages} {060401} (\bibinfo {year}
  {2013})}\BibitemShut {NoStop}%
\bibitem [{SI()}]{SI}%
  \BibitemOpen
  \href@noop {} {\emph {\bibinfo {title} {See Supplemental Materials for details.}}}\BibitemShut {Stop}%
\bibitem [{\citenamefont {Freund}(1990)}]{Freund}%
  \BibitemOpen
  \bibfield  {author} {\bibinfo {author} {\bibfnamefont {L.~B.}\ \bibnamefont
  {Freund}},\ }\href@noop {} {\emph {\bibinfo {title} {Dynamic Fracture
  Mechanics}}}\ (\bibinfo  {publisher} {Cambridge University Press},\ \bibinfo
  {address} {Cambridge},\ \bibinfo {year} {1990})\BibitemShut {NoStop}%
\bibitem [{\citenamefont {Bonamy}\ and\ \citenamefont
  {Bouchaud}(2011)}]{bonamy2011failure}%
  \BibitemOpen
  \bibfield  {author} {\bibinfo {author} {\bibfnamefont {D.}~\bibnamefont
  {Bonamy}}\ and\ \bibinfo {author} {\bibfnamefont {E.}~\bibnamefont
  {Bouchaud}},\ }\bibfield  {title} {\bibinfo {title} {Failure of heterogeneous
  materials: A dynamic phase transition?},\ }\href@noop {} {\bibfield
  {journal} {\bibinfo  {journal} {Physics Reports}\ }\textbf {\bibinfo {volume}
  {498}},\ \bibinfo {pages} {1} (\bibinfo {year} {2011})}\BibitemShut {NoStop}%
\bibitem [{\citenamefont {Bouchbinder}\ \emph {et~al.}(2008)\citenamefont
  {Bouchbinder}, \citenamefont {Livne},\ and\ \citenamefont
  {Fineberg}}]{bouchbinder.08a}%
  \BibitemOpen
  \bibfield  {author} {\bibinfo {author} {\bibfnamefont {E.}~\bibnamefont
  {Bouchbinder}}, \bibinfo {author} {\bibfnamefont {A.}~\bibnamefont {Livne}},\
  and\ \bibinfo {author} {\bibfnamefont {J.}~\bibnamefont {Fineberg}},\
  }\bibfield  {title} {\bibinfo {title} {Weakly nonlinear theory of dynamic
  fracture},\ }\href@noop {} {\bibfield  {journal} {\bibinfo  {journal}
  {Physical Review Letters}\ }\textbf {\bibinfo {volume} {101}},\ \bibinfo
  {pages} {264302} (\bibinfo {year} {2008})}\BibitemShut {NoStop}%
\bibitem [{\citenamefont {Goldman}\ \emph {et~al.}(2012)\citenamefont
  {Goldman}, \citenamefont {Harpaz}, \citenamefont {Bouchbinder},\ and\
  \citenamefont {Fineberg}}]{goldman2012}%
  \BibitemOpen
  \bibfield  {author} {\bibinfo {author} {\bibfnamefont {T.}~\bibnamefont
  {Goldman}}, \bibinfo {author} {\bibfnamefont {R.}~\bibnamefont {Harpaz}},
  \bibinfo {author} {\bibfnamefont {E.}~\bibnamefont {Bouchbinder}},\ and\
  \bibinfo {author} {\bibfnamefont {J.}~\bibnamefont {Fineberg}},\ }\bibfield
  {title} {\bibinfo {title} {Intrinsic nonlinear scale governs oscillations in
  rapid fracture},\ }\href@noop {} {\bibfield  {journal} {\bibinfo  {journal}
  {Physical Review Letters}\ }\textbf {\bibinfo {volume} {108}},\ \bibinfo
  {pages} {104303} (\bibinfo {year} {2012})}\BibitemShut {NoStop}%
\bibitem [{\citenamefont {Bouchbinder}\ \emph {et~al.}(2014)\citenamefont
  {Bouchbinder}, \citenamefont {Goldman},\ and\ \citenamefont
  {Fineberg}}]{bouchbinder.14}%
  \BibitemOpen
  \bibfield  {author} {\bibinfo {author} {\bibfnamefont {E.}~\bibnamefont
  {Bouchbinder}}, \bibinfo {author} {\bibfnamefont {T.}~\bibnamefont
  {Goldman}},\ and\ \bibinfo {author} {\bibfnamefont {J.}~\bibnamefont
  {Fineberg}},\ }\bibfield  {title} {\bibinfo {title} {The dynamics of rapid
  fracture: instabilities, nonlinearities and length scales},\ }\href@noop {}
  {\bibfield  {journal} {\bibinfo  {journal} {Reports on Progress in Physics}\
  }\textbf {\bibinfo {volume} {77}},\ \bibinfo {pages} {046501} (\bibinfo
  {year} {2014})}\BibitemShut {NoStop}%
\end{thebibliography}

\end{document}